\documentclass[twocolumn,amsmath,aps,nofootinbib,superscriptaddress]{revtex4}

\usepackage{amssymb}
\usepackage{amsmath}
\usepackage{epsfig}
\usepackage{subfigure}
\usepackage{mathrsfs}
\usepackage{longtable}
\usepackage[usenames,dvipsnames]{xcolor}

\usepackage{mathtools}
\usepackage{relsize}

\newcommand{\be}{\begin{equation}}
\newcommand{\ee}{\end{equation}}
\newcommand{\bea}{\begin{align}}
\newcommand{\eea}{\end{align}}

\newcommand{\unit}[1]{\ensuremath{\, \mathrm{#1}}}

\newcommand{\gb}{\gamma_{\rm b}}

\newcommand{\chimin}{\langle \chi \rangle}

\begin{document}

\title{Cosmology of Axions and Moduli: A Dynamical Systems Approach}

\author{David J. E. Marsh}
\email{d.marsh1@physics.ox.ac.uk}
\affiliation{Rudolf Peierls Centre for Theoretical Physics, University of Oxford, 1 Keble Road, Oxford, OX1 3NP, UK}
\author{Ewan R. M. Tarrant}
\email{ppxet@nottingham.ac.uk}
\affiliation{School of Physics and Astronomy, University of Nottingham, University Park, Nottingham, NG7 2RD, UK}
\author{Edmund J. Copeland}
\email{ed.copeland@nottingham.ac.uk}
\affiliation{School of Physics and Astronomy, University of Nottingham, University Park, Nottingham, NG7 2RD, UK}
\author{Pedro G. Ferreira}
\email{p.ferreira1@physics.ox.ac.uk}
\affiliation{Astrophysics, University of Oxford, DWB, Keble Road, Oxford, OX1 3RH, UK}

\date{\today}

 %---------------------- ABSTRACT -------------------------

\begin{abstract}

This paper is concerned with string cosmology and the dynamics of multiple scalar fields in potentials that can become negative, and their features as (Early) Dark Energy models. Our point of departure is the ``String Axiverse'', a scenario that motivates the existence of cosmologically light axion fields as a generic consequence of string theory. We couple such an axion to its corresponding modulus. We give a detailed presentation of the rich cosmology of such a model, ranging from the setting of initial conditions on the fields during inflation, to the asymptotic future. We present some simplifying assumptions based on the fixing of the axion decay constant $f_a$, and on the effective field theory when the modulus trajectory is adiabatic, and find the conditions under which these assumptions break down. As a by-product of our analysis, we find that relaxing the assumption of fixed $f_a$ leads to the appearance of a new meta-stable de-Sitter region for the modulus without the need for uplifting by an additional constant. A dynamical systems analysis reveals the existence of many fixed point attractors, repellers and saddle points, which we analyse in detail. We also provide geometric interpretations of the phase space. The fixed points can be used to bound the couplings in the model. A systematic scan of certain regions of parameter space reveals that the future evolution of the universe in this model can be rich, containing multiple epochs of accelerated expansion. 

%%%%%%%%%

\end{abstract}

\maketitle
% -------------------------------------------------------------------------

%\setlength{\parskip}{1cm plus4mm minus3mm}

\section{Introduction}
\label{intro}

String cosmology \cite{copeland1994,lidsey1999,mcallister2008,burgess2011} has been undergoing a renaissance over the last decade, due in part to our increased understanding of the landscape \cite{polchinski2006,bousso2006}, advances in moduli stabilisation and string model building \cite{kachru2003,conlon2006,cicoli2011a} and the continued successes of precision cosmology \cite{7yearWMAP}. Theory and experiment have found fruitful harmony in two main areas: the early time accelerated expansion of the universe and generation of primordial density perturbations during inflation \cite{guth1981,Starobinsky:1980te,gasperini1993,copeland1997,kachru2003b,cicoli2011b,dimopoulos2012} and in the study of the current epoch of cosmological acceleration \cite{riess1998,perlmutter1999} through theories of dark energy and quintessence \cite{copeland2006}. Recently, the ``String Axiverse'' \cite{axiverse2009} has motivated the study of ultra-light axion fields that evolve on cosmological time scales \cite{acharya2010a,marsh2010,arvanitaki2010,dubovsky2010,marsh2011b,kodama2011,higaki2011}: one should ask whether the axion dynamics in this scenario can have other interesting or adverse effects on string cosmology and model building.

Axions have been ubiquitous in theoretical physics since they were first proposed to solve the strong CP problem \cite{pecceiquinn1977}. Ever since this time, they have presented many problems and possibilities to cosmologists \cite{dine1981, preskill1983, steinhardt1983, turner1983, abbott1983, dine1983,ipser1983,turner1986,berezhiani1991,berezhiani1992,jmr1992,
banks1996,fox2004,hannestad2005,visinelli2009,hannestad2010} (for reviews of axion cosmology, see \cite{sikivie2008,sikivie2010}). Indeed, axions are the major contender with Weakly Interacting Massive Particles (WIMPs) as a main constituent of the Dark Matter (DM). If both axions and WIMPs exist, it is natural that they should share the DM burden \cite{aguirre2004}.

In this paper we systematically explore the parameter space and dynamics of the coupled axion-modulus system originally proposed in \cite{marsh2011}, which is a phenomenological extension of the axiverse to include effects on moduli. The energy scales of this model cannot be argued so model independently and elegantly to give rise to interesting cosmology as in the axiverse alone. However we take a phenomenological approach and use it to ask the questions: can cosmological axion dynamics lead to cosmological modulus dynamics? What are these dynamics? For what parameter values does interesting phenomenology occur? It is the purpose of this paper to answer  these questions.

\renewcommand*\arraystretch{1.1}

\begin{center}
\begin{table*}[!htp]
\hfill{}
\begin{tabular}{c|l|c}
\hline
\hline
Symbol			&	 Meaning  & Reference Eq. \\
\hline
$f_a$     			&      axion decay constant   &  (\ref{eqn:axionLagrangain})\\
\hline
$\Lambda_a$ 		& 	axion potential energy scale  &  (\ref{eqn:axionLagrangain})\\
\hline
$\mu$		& 	energy scale of non--perturbative physics for axion potential  & (\ref{eqn:scales}) \\
\hline
$B\,,D$		&  modulus potential parameters	  & (\ref{eqn:potential}) \\
\hline
$\rho_\Lambda$		&  cosmological constant energy density	&  (\ref{eqn:potential})\\
\hline
$\Lambda$		&  value of the true vacuum energy	&  (\ref{eqn:vacuum_cc})\\
\hline
$\phi$		&  axion field	& (\ref{eqn:axionLagrangain_can}) \\
\hline
$\chi$		& 	 modulus field	&  (\ref{eqn:potential})\\
\hline
$C$		& 	axion--modulus coupling constant  &  (\ref{eqn:potential})\\
\hline
$\gb$		&  baryotropic fluid equation of state	&  (\ref{eqn:axionModuliEoM})\\
\hline
$x\,,y\,,z\,,r\,,s\,,t$		&  autonomous system variables	&  (\ref{eqn:autonomousSystem})\\
\hline
$N_{\rm efd}$		&  number of $e$--folds from beginning of model evolution until end of fluid domination	& -- \\
\hline
$\mathcal{N}_{ae}$		&  number of periods of accelerated expansion	& -- \\
\hline
$\Omega_e$		&   early dark energy (EDE) density    &   (\ref{eqn:omegae}) 	\\
\hline
\hline
$M$		&  $=\mu^2/f_a$ (axion mass scale)	&   (\ref{eqn:potential}) \\
\hline
$\beta$		& $=\sqrt{3M^2/D}$	  &    (\ref{eqn:beta})\\
\hline
$\tilde{\phi}$		&  $=\sqrt{6}/\beta$ (critical value of $\phi$ above which modulus is destabilised)	& (\ref{eqn:minbound}) \\
\hline
$\omega$		& $=\sqrt{B/D}$	    &   (\ref{eqn:chimin}) \\
\hline
$\zeta$		&   $=B\rho_\Lambda/D^2=t^2z^2/r^4$  (vacuum constraint). $\zeta>(<)1/4$ gives dS (AdS)	&  (\ref{eqn:defZeta}) \\
\hline
\end{tabular}
\hfill{}
\caption{Symbols used in this paper.}
\label{tab:symbols}
\end{table*}
\end{center}

When dimensionally reduced and considered at low energies, string theory furnishes us with extra degrees of freedom, in addition to the $\Lambda$CDM concordance model ingredients of General Relativity, cold dark matter (CDM), and the standard model of particle physics. In fact, cosmologically relevant axions and moduli are \emph{the} generic prediction of string/M-theory \cite{acharya2012}. These extra degrees of freedom can be viewed as a blessing or a curse. Extra scalar fields with appropriately fine tuned potentials in the early universe are useful for inflationary model building, but they also lead to the cosmological moduli problem and must be properly stabilised. Scalar fields can also serve as dark matter, or dark energy. The simplest 6-parameter version of concordance $\Lambda$CDM cosmology may soon be observationally extended with the detection of neutrino mass and mass splittings \cite{bernardis2009}. There are also observational hints from small-scale CMB experiments \cite{dunkley2010,keisler2011} that other relativistic degrees of freedom can already be seen in cosmology. The possibility that this may be string theory related, for example through models of Early Dark Energy \cite{wetterich2004,doran2006}, is an exciting one. Independently of string theory, a multi-component dark sector is natural in, and may be strong evidence for, the top-down or anthropic view of cosmology \cite{aguirre2004}. 

The landscape of string theory vacua can also be seen as a blessing or a curse: it hampers the exact predictivity of string theory, but it may yet explain the smallness of the cosmological constant by providing a high enough density of possible vacua ``near'' to a phenomenologically acceptable one~\cite{bousso2006}, and through eternal inflation provide a mechanism for the scanning. In the model we will study, the overall value of the vacuum energy is a free parameter; it was shown in \cite{marsh2011} that one can have this negative, so that the universe today is rolling towards collapse. Harlow et al \cite{harlow2012,susskind2012} have argued that the existence of terminal vacua is necessary for the existence of a global arrow of time in eternal inflation, and studying their phenomenology therefore seems pertinent. In all cases what we gain from string cosmology is the potential to ask deeper questions when models are embedded in a UV-complete theory.  

In our view, another bright side of the string landscape comes from looking for the ``why not?'' features. The axiverse is one such generic feature. Along with the moduli of the landscape, we also get axions. We will save the details of this scenario for later, for now all we need say is that many of these axions should remain light, and thus cosmologically active. Through their effects on structure formation \cite{marsh2010} it will be possible with next-generation cosmological observations to constrain the existence of these fields as a component of the dark sector energy density at percent-level accuracy \cite{marsh2011b}. The axiverse raises the possibility that string cosmology may be active at late times and that we may be able to observe it. This has been called the ``Low-Energy Frontier of Particle Physics'' \cite{jaeckel2010b}. If we can detect axions with high precision using cosmology, might we also detect changes in their evolution caused by the moduli?

As the simplest models start to be constrained, we can begin to explore them more deeply. Naturally, one goes from assuming that just one scalar field is active at late times, to assuming that many are \cite{dienes2011a,dienes2011b,dienes2012}. Already in inflation, thought of as embedded in string theory, the dominant paradigm is of an inflationary direction in a multi-dimensional field space (e.g. \cite{bond2007,blanco-pillado2009,cicoli2012}). It should also be true that the light axions of the axiverse exist as a flat direction in some much larger field space of their brother axions and sister moduli. To look simply at many axion fields, and ignore their partner moduli is the most conservative option: the axiverse should be general enough to exist independently of the mechanism for moduli stabilisation. But we will show here that there are cases where one cannot ignore the effects of the moduli, and so we explore what these effects might be within a certain parameterisation. 

This paper is organised as follows: In Section~(\ref{model}) we recall the coupled axion-modulus system introduced in \cite{marsh2011} and comment on how the initial conditions appropriate for acceptable late time phenomenology might arise. In Section~(\ref{potential}) we analyse the basic features of cosmology in this model by looking at the scalar potential, before presenting a systematic analysis of the dynamics using a dynamical systems approach in Section~(\ref{phasespace}).  We discuss our findings in Section~(\ref{discussion}), and conclude in Section~(\ref{conclusions}) 

Our detailed presentation of the rich cosmology of our model requires the use of many equations and symbols. As a guide to the reader, in Table~(\ref{tab:symbols})  we summarize the key symbols and the equations where they are defined or first used.

% -------------------------------------------------------------------------
\section{The Model}\label{model}

\subsection*{The Axiverse}

All types of string theory and M-theory contain multiple axion fields \cite{witten1984,witten2006}. These axions arise when anti-symmetric tensor fields are compactified on closed cycles; the axion is the Kaluza-Klein zero mode, and appears in the gauge kinetic function. Axions then acquire a potential when non-perturbative physics is turned on on the cycle, for example from wrapped D-branes, or from world sheet or gauge theory instantons. The axion is a pseudo-Nambu-Goldstone boson (PNGB) of a spontaneously broken global symmetry. Many extensions of the standard model of particle physics also contain more generic PNGBs \cite{hill1988}.

The low energy four dimensional Lagrangian for an axion, $\theta$, with periodic potential $U(\theta)$ is:
\begin{equation}
\mathcal{L}=\frac{f_a^2}{2}(\partial \theta)^2-\Lambda_a^4 U(\theta)\,.
\label{eqn:axionLagrangain}
\end{equation}
The two scales in this Largrangian, the decay constant $f_a$, and potential energy scale $\Lambda_a$, both depend upon the action, $S$, of the non-perturbative physics on the corresponding cycle in the following way:
\begin{align}
f_a&\sim \frac{M_{pl}}{S}\,, \nonumber \\
\Lambda_a^4 &= \mu^4 e^{-S}\,. \nonumber \\
\label{eqn:scales}
\end{align}
Here $M_{pl}$ is the (reduced) Planck mass and $\mu$ sets the scale of non-perturbative physics, for example the QCD scale or, in string theory, the geometric mean of the supersymmetry (SUSY) breaking scale, $M_{\rm SUSY}$ and the Planck scale, $\sqrt{M_{pl} M_{\rm SUSY}}$\footnote{In \cite{marsh2011} this mean was misquoted. A forthcoming erratum will correct this. The argument and scales in that work should follow from a higher scale of SUSY breaking}\cite{axiverse2009}. The action in turn depends on the size of the cycle, and herein we find the axiverse mechanism for light axions. Although $f_a$ should be fixed at some high scale $f_a \sim 10^{16}\unit{GeV}$ \cite{witten2006}, small variations in the sizes of cycles and the exponential sensitivity of $\Lambda_a$ on $S$ means that the axion mass should distribute roughly evenly on a logarithmic scale, leading to some ultra-light, stable axions for the cosmologist to play with. 

We canonically normalise $\phi = f_a \theta$, and expand the potential around the minimum:

\begin{align}
\mathcal{L} = \frac{1}{2}(\partial \phi)^2 &- \frac{1}{2}m_a^2\phi^2\,, \nonumber \\
m_a^2 &= \frac{\Lambda_a^4}{f_a^2}\,. \nonumber \\
\label{eqn:axionLagrangain_can}
\end{align}

% -------------------------------------------------------------------------
\subsection*{The Axiverse Is More Than Just Axions}

As with everything in string theory, our low energy ``constants'', such as $m_a$, are not really constant at all, but depend upon moduli. In this case the modulus of interest is that controlling the area of the cycle giving us the axion. $S$ depends on this area, and so we can choose to identify $S=C\chi$ for some modulus field $\chi$, and coupling $C$. Eq.~(\ref{eqn:scales}) then implies that \emph{axions and moduli are coupled}. 

This modulus must be stabilised non-perturbatively \cite{book:becker}. Typical potentials are sums of exponentials (see, for example, \cite{kachru2003,conlon2006,conlon2010,acharya2010a,cicoli2011a}). These considerations led, in \cite{marsh2011}, to the following potential being studied for a coupled axion-modulus system\footnote{ (i) The general form of the potential we study fits into the class of models of ``Generalized Assisted Inflation'' \cite{liddle1998,copeland1999,hartong2006}, though we mainly emphasise its use for quintessence, rather than inflationary, purposes. (ii) At large values of $\chi$, loop effects will eventually cause the potential to rise again, as in \cite{cicoli2011a,cicoli2011c}. We do not consider such contributions. Our conclusions only depend on having a sufficiently long, flat region of the potential before these effects kick in, but we do not compute the scales of parameters necessary for this.}:
\begin{equation}
V(\phi,\chi) = B e^{-2C\chi}-D e^{-C\chi}+\frac{1}{2}M^2 e^{-C\chi}\phi^2 +\rho_\Lambda\,,
\label{eqn:potential}
\end{equation}
where $\rho_\Lambda$ is the cosmological constant, added \emph{arbitrarily} in this model so that whatever value the potential takes today can be made consistent with observations. We also have $M^2 = \mu^4/f_a^2$. The total Lagrangian is then of the form:
\begin{equation}
\mathcal{L}= \frac{1}{2}(\partial \phi)^2 + \frac{1}{2}(\partial \chi)^2 - V(\phi,\chi)\,.
\end{equation}

There is one important caveat to this picture: by taking $f_a$ fixed we are implicitly assuming small modulus variations. This greatly simplifies our system, since if we allowed for the variation of $f_a(\chi)$ this would change the canonical normalisation of the axion kinetic terms, and introduce kinetic mixing between the $\phi$ and $\chi$ fields. This effect could introduce new phenomenology in extreme trajectories with large $\Delta \chi/\chi$, for example the possibility of chaotic behaviour, but we defer study of this to a future work\footnote{We thank John March-Russell for pointing this fact out to us.}. We make some comments on the effect on the potential in Section~(\ref{discussion}).

We assume that the axion and modulus fields evolve in a spatially flat Friedmann-Lemaitre-Robertson-Walker (FLRW) background containing the bare cosmological constant, $\rho_\Lambda$, and a single fluid with baryotropic equation of state $P_{\rm b}=(\gb-1)\rho_{\rm b}$, where $\gb$ is a constant, $0\le\gb\le2$. For radiation $\gb=4/3$ or for dust CDM $\gb=1$. In standard cosmic time the evolution equations are:
\begin{eqnarray}
	\ddot{\phi}&=&-3H\dot{\phi}-\frac{\partial V}{\partial \phi} \,, \nonumber \\
	\ddot{\chi}&=&-3H\dot{\chi}- \frac{\partial V}{\partial \phi}  \,, \nonumber \\
	\dot{\rho}_{\rm b}&=&-3H\rho_{\rm b}\gb \,, \nonumber \\
	\dot{H}&=&-\frac{1}{2}[\rho_{\rm b}\gamma_{\rm b}+\dot{\phi}^2+\dot{\chi}^2] \,.
	\label{eqn:axionModuliEoM}	
\end{eqnarray}
These are subject to the Friedmann constraint
\begin{equation}
	3H^{2} = \rho_{b} + \rho_{\Lambda} + \rho_{\phi} + \rho_{\chi} \,,
	\label{eqn:Friedmann}	
\end{equation}
where we have assumed that we can define the following distinct densities and pressures:
\begin{eqnarray}
	\rho_{\phi} &=& \frac{1}{2}\dot{\phi}^{2} + U(\phi,\chi) \,, \nonumber \\
	P_{\phi}      &=&  \frac{1}{2}\dot{\phi}^{2} - U(\phi,\chi)  \,, \nonumber \\
	\rho_{\chi} &=&  \frac{1}{2}\dot{\chi}^{2} + V_{B}(\chi) - V_{D}(\chi)  \,, \nonumber \\
	P_{\chi}      &=&  \frac{1}{2}\dot{\chi}^{2} -  V_{B}(\chi) + V_{D}(\chi) \,.
	\label{eqn:densityPressureAxionModulus}	
\end{eqnarray}
by splitting the potential as:
\begin{equation}
	V(\phi,\chi) = V_{B}(\chi) - V_{D}(\chi) + U(\phi,\chi) +\rho_\Lambda \,,
	\label{eqn:potentialVsplit}	
\end{equation}
where
\begin{eqnarray}
	 V_{B}(\chi)   &=&   Be^{-2C\chi}  \,,  \nonumber \\
	 V_{D}(\chi)   &=&  De^{-C\chi}   \,, \nonumber \\
	 U(\phi,\chi)  &=&  \frac{1}{2}e^{-C\chi}M^{2}\phi^{2}   \,.
	\label{eqn:potentials}	
\end{eqnarray}

We note that the split in these cases between dark matter and dark energy is somewhat arbitrary \cite{kunz2007}. When scalar fields begin oscillating, they redshift and cluster as dark matter, with an individual equation of state $w_{\phi\chi} \rightarrow 0$, but the modulus term also has a vev, behaving as a cosmological constant. Slow roll, and the inclusion of $\rho_\Lambda$ muddy the waters further. Also, it is completely arbitrary in the presence of coupling to place $U(\phi,\chi)$ in $\rho_\phi$.

In the axiverse as presented in \cite{axiverse2009} all moduli were assumed to be absolutely stabilised. The different sizes at which they were stabilised led to the different masses for the axions. Some were stabilised at larger values than others in order to make some axions light, but the differences are not hierarchical. Assuming absolute stability implied that the moduli were heavy, and lived in their global minimum. The axiverse has been concretely realised in the moduli stabilisation scheme of \cite{acharya2010a}. In the scheme of \cite{cicoli2011a}, moduli were stabilised at hierarchically different values, which allowed some moduli to remain very light. In this picture the axion phenomenology is not considered, and they are set to their vacuum values at zero. This is perfectly well justified even for light axions if all we are concerned with is the existence of a stable minimum for the moduli and in calculating their masses at this minimum, but if the axion evolution is our focus, then their possible effects on the moduli cannot be ignored. In both \cite{acharya2010a} and \cite{cicoli2011a} only a handful of fields were considered, not the hundreds motivated in the axiverse. Here we take inspiration from the success of these models, and apply to it the spirit of optimism of the axiverse to look for phenomenology in a larger arena of possibilities.

% -------------------------------------------------------------------------

\subsection*{Cosmic Overview}
\label{inflation}

Here we give an overview of a scenario that may lead to the realisation of the initial conditions appropriate to our model, and the picture of cosmic history that emerges.

We will assume that the universe begins in an eternally inflating de-Sitter (dS) false vacuum. This vacuum decays via tunnelling and bubble nucleation \cite{coleman1980} into the standard phase of slow-roll inflation required to generate the primordial power spectrum. It must also be assumed that initial conditions on the axion and modulus fields are laid down \emph{prior} to inflation. After slow-roll inflation ends, the inflaton decays and reheats the universe. We will consider the toy model of a post-inflation universe consisting only of matter, radiation, and the axion and modulus field condensates contributing a dark sector energy density as described in \cite{marsh2011}.

In \cite{marsh2011}, initial conditions were such that the modulus began at a large value, $\chi_i$ and the axion mass $m_a = M^2 e^{- C \chi}$ was cosmologically light, $m_a = \mathcal{O}(1-10^{10})H_0$. This modulus initial condition was \emph{not} at the local minimum of the potential, $\chimin (\phi)$. The axion initial condition, $\phi_i$ is set at the Peccei-Quinn (PQ) phase transition by spontaneous symmetry breaking \cite{pecceiquinn1977,book:kolb_and_turner}. There are two logical possibilities for modulus evolution: there is a local modulus minimum at $\phi_i$, or there is not. If there is a minimum, and the modulus is sufficiently heavy to overcome Hubble friction, it will, like the inflaton before it, roll to the \emph{local} minimum, while the light axion frozen at $\phi_i$ prevents it from reaching the global minimum. If there is no local minimum, then the modulus will roll to yet larger values until stopped by Hubble friction, only decaying to a local minimum once one exists (see e.g. \cite{conlon2008})\footnote{These considerations are basically a statement of the Brustein-Steinhardt problem \cite{brustein1993} for this model. Related issues are discussed in \cite{skinner2003}.}. Under these conditions, although the modulus and axion masses at the global minimum could be large (e.g. string/Planck scale), interactions instead freeze the modulus either in its local minimum, or at $\chi>\chi_i$. In both cases, the modulus \emph{must} evolve with the axion.

For these initial conditions to be possible the only requirement is that the PQ phase transition and the switching on of the appropriate instanton effects, which create the axion condensate and form the coupled potential, happen \emph{before the modulus finds its minimum}. Certainly, during slow-roll inflation there are scalar fields yet to find their minima: the inflaton itself is one such field. It is not unreasonable to assume that there are other moduli present that also exist away from their minima. Indeed, this is the case in any model of multi--field inflation and is the string interpretation of any quintessence model. This is also the expectation for the post-inflation, pre-hot-big-bang phase in string cosmology, where the post-inflation universe is dominated by the yet-to-decay, matter-like moduli \cite{banks1996,acharya2009,acharya2010a}.

The requirement that inflation occurs after the PQ phase transition, and that the reheat temperature does not restore the PQ symmetry, is generic to almost all models with axions as it is required to avoid a cosmological abundance of disastrous domain walls and the like \cite{book:kolb_and_turner}. In addition, string axion models require a low energy scale of inflation \cite{axiverse2009}, in part to avoid overproduction of axion isocurvature perturbations.

After the radiation and matter dominated phases end, the next stage in the evolution of the universe again has a number of possibilities depending on the axion and modulus fields. The additional cosmological constant in the potential, $\rho_\Lambda$, can be regarded as the usual left over contribution to the vacuum energy. It has contributions which reduce it over time after spontaneous symmetry breaking (e.g. at the electroweak transition), positive contributions from dS stabilised moduli and vacuum fluctuations of quantum fields, and negative contributions from AdS stabilised moduli (see e.g. \cite{bousso2006}). The value of this constant relative to the potential minimum in the axion and modulus fields determines the fate of the universe.

If the axion and modulus fields are heavy and their field values are small, such that they are oscillating about, or slowly rolling into, their global minimum during the present epoch, then the scenario will be much like any other quintessence or axion dark matter scenario. The value of the total cosmological constant in the bottom the potential must be small, and of the correct magnitude to account for the observed accelerated expansion of the universe. However, if the fields are light enough and their initial values large enough that they are on the plateau of the potential, then the phenomenology can be quite different. Here, the potential energy of the axion and modulus are a small contribution, and the current accelerated expansion will be driven almost entirely by $\rho_\Lambda$, as was the case in \cite{marsh2011}.

However, as also described in \cite{marsh2011}, the axion-modulus system is only quasi-stable: eventually axion oscillations will decay and the modulus will find the global minimum. Depending on the initial conditions and the parameters in the potential it is possible to arrange for an acceptable cosmology where the vacuum energy at the global minimum is either positive, negative, or indeed zero. If the vacuum energy is negative then the decay of the modulus will trigger rapid cosmological collapse, rather than life in a stable AdS state \cite{coleman1980,banks2002,steinhardt2002}. In the case where it is positive, then a scenario such as explored in \cite{amin2011} will ensue.

% -------------------------------------------------------------------------
\section{Axion-Modulus Dynamics and the Coupled Potential}
\label{potential}

Before we begin our detailed dynamical systems analysis in Section~(\ref{phasespace}), we aim to give here some basic intuition about the types of phenomena possible in a cosmology with coupled scalar fields and an arbitrary vacuum energy. In particular, we give examples of phenomena not explored previously in \cite{marsh2011}. The examples use arbitrary values of the parameters and are for illustration only.

% -------------------------------------------------------------------------
\subsection{Local Minima, the Adiabatic Trajectory, and an Effective Potential}

The form of the potential is such that there is just one minimum, when the axion is at zero. However, for light axions, most of cosmic history~\footnote{When viewed in redshift space, where we have cosmic coincidence for things occurring within $z=1$ of us, which is of order billions of years. The coincidence problem is in fact much more of a problem \emph{forwards} in time: why are we not Boltzmann brains in thermal de-Sitter space?} is spent away from this global minimum. The form of the coupling between axion and modulus then means that the moduli, too, will live away from their global minimum and consequently \emph{moduli must evolve during the course of cosmic history}. 

\begin{figure}
\centering
\includegraphics[scale=0.78]{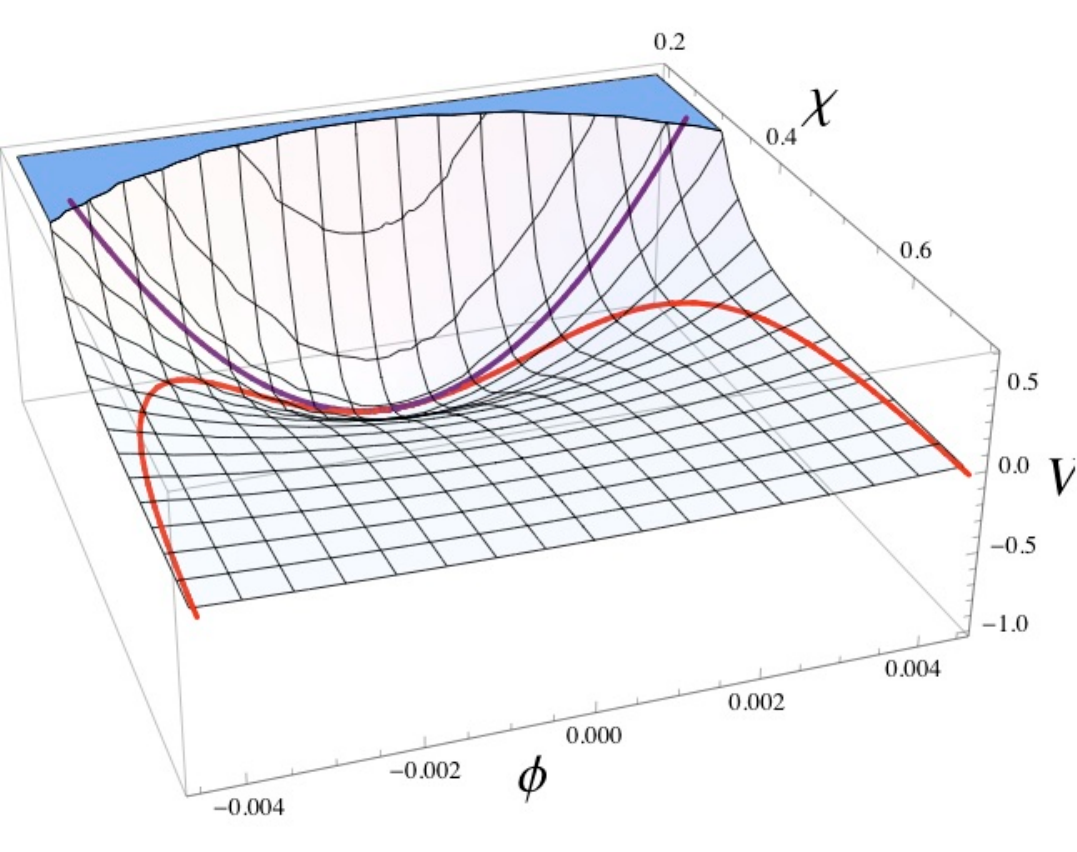}
\caption{The potential of Eq.~(\ref{eqn:potential}) near the global minimum, for arbitrary parameters. In red, the adiabatic trajectory: the modulus minimum as a function of $\phi$ (Eq.~(\ref{eqn:chimin})). When $\phi$ takes large values (defined by Eq.~(\ref{eqn:minbound})), the minimum at finite $\chi$ is destroyed. In purple, the naive trajectory: the axion potential at fixed $\chi=\chimin (0)$. For a heavy modulus the adiabatic trajectory will be followed, which is shallower near the minimum than the naive trajectory.}
\label{fig:potential1}
\end{figure}

We find that the local minimum in the modulus direction as a function of the axion background is:
\begin{equation}
\chimin (\phi) = -\frac{1}{C} \ln \left[ \frac{1}{2 \omega^2}\left( 1-\frac{\beta^2}{6}\phi^2 \right) \right]\,,
\label{eqn:chimin}
\end{equation}
where
\begin{equation}
	\beta=\sqrt{\frac{3M^2}{D}}\,.
	\label{eqn:beta}	
\end{equation}
and $\omega^2 = B/D$. We plot this trajectory on the potential surface in Fig.~(\ref{fig:potential1}).

The existence of the global minimum at positive modulus translates into the bound: $\omega >\frac{1}{\sqrt{2}}$. We also find that there is no local modulus minimum for large axion field values:
\begin{equation}
\phi> \frac{\sqrt{6}}{\beta} \equiv \tilde{\phi}\quad \Rightarrow \text{no modulus minimum}\,.
\label{eqn:minbound}
\end{equation}
That is to say: \emph{for large axion field values, the corresponding modulus will become destabilised}\footnote{The same is true when the variation of $f_a$ with $\chi$ is taken into account, but the condition must be found numerically.}. The disappearance of the minimum at large axion field values is precisely the appearance of the ``valley walls'' in the potential, as described in \cite{marsh2011}.

When the axion has a periodic potential, canonically of the form $U(\theta)=1-\cos (\theta)$, then the axion has a maximum field value at $\theta=\pi$. Such a periodic field can spoil the local modulus minimum when:
\begin{equation}
\frac{\mu^4}{D}>\frac{1}{2}\,.
\label{eqn:shift_bound}
\end{equation}

Eqs.~(\ref{eqn:minbound}, \ref{eqn:shift_bound}) show that if the natural scales in the axion potential (either $M^2$ or $\mu^4$) arising from non-perturbative physics, are of the same order or slightly larger than the natural scales in the modulus potential (in this case $D$), which are also non-perturbative, then destabilisation can occur even for small field values. Whether or not this mild hierarchy of scales occurs in actual models of moduli stabilisation is not the subject of this work, but we see no \emph{a priori} reason why it should not be possible.

If the axion initial conditions are such that field values are large then the corresponding modulus has no potential minimum in the early universe. If the axion undergoes monodromy \cite{silverstein2008,panda2010,amin2011} the shift symmetry is broken and large field values are natural. For smaller axion field values this condition can still be satisfied for sufficiently large $\beta^2$, or if the bound of Eq.~(\ref{eqn:shift_bound}) is satisfied. 

In the rest of this section we will be concerned with situations where a local minimum for $\chi$ \emph{does} exist. In this case where there is a local modulus minimum, there is still interesting physics caused by the axion background. If the modulus begins life at its local minimum in the frozen axion background then the fractional change in the modulus field during axion evolution from $\phi=\phi_i$ to $\phi=0$ is:
\begin{equation}
\frac{\Delta \chi}{\chimin (0)} = \frac{|\Delta f_a |}{f_{a,i}}=\frac{\ln \left( 1-\left( \frac{\phi_i}{\tilde{\phi}} \right)^2 \right)}{\ln \left( \frac{1}{2 \omega^2} \right)}\,.
\end{equation}
This ratio blows up when $\phi_i=\tilde{\phi}$, where the modulus is destabilised and the local minimum is at $f_{a,i} = 0$. However, it remains $\mathcal{O}(1)$ for $\phi<\tilde{\phi} \sqrt{1-\frac{1}{2 \omega^2}}$. As we will see, $\omega$ does not appear in our dynamical system analysis and so can be picked arbitrarily large (corresponding to stabilising the modulus at larger and larger values) and these results can be made insensitive to the approximation that $f_a$ is fixed. For the consistency of our assumption that PQ symmetry is broken before inflation, we must have $f_a$ larger than the inflationary energy scale, and it must remain large enough that the symmetry is never restored. Two comments are in order here. Firstly, the modulus will never roll out to truly infinite values because of Hubble friction. Secondly, however, we may in general expect trajectories that go from a destabilised region on the plateau of the potential into the global minimum to require some fine tuning in order not to break our assumption of fixed $f_a$.  We will comment more on this later.

The ratio of modulus mass at the start and end of this trajectory is:
\begin{equation}
\frac{m_\chi (\chimin (\phi_i))}{m_\chi (\chimin (0))} = 1-\left( \frac{\phi_i}{\tilde{\phi}} \right)^2\,.
\label{eqn:massratio}
\end{equation}
This ratio can become small as the bound of Eq.~(\ref{eqn:minbound}) becomes saturated. When this bound is saturated, or nearly saturated, and if other corrections to the modulus mass are small\footnote{For example, suitable decoupling occurs in the scenario of \cite{cicoli2011a}.}, then an effective field theory obtained by integrating out the modulus based on its mass at the global minimum may fail. The potential is anharmonic, and so if the modulus is displaced far from its local minimum the mass will not be given by this formula.

The ratio of Eq.~(\ref{eqn:massratio}) occurs also for the axion mass along this trajectory, implying that in such a situation the fields cannot change their relative masses during the course of their evolution. If the modulus is heavier at the global minimum, it will also be heavier in any local minimum. This allows for consistency of the assumption above: if the modulus minimises first it should be a good approximation in this case to consider the trajectory as being $\chimin (\phi)$. We will call this trajectory the \emph{adiabatic trajectory}, i.e. the one that the modulus follows if it is always heavy enough to return to equilibrium sufficiently quickly as the axion rolls.

%%%
\begin{figure*}[htp]
\centering
$\begin{array}{@{\hspace{0in}}l@{\hspace{0in}}l}
\includegraphics[scale=0.65]{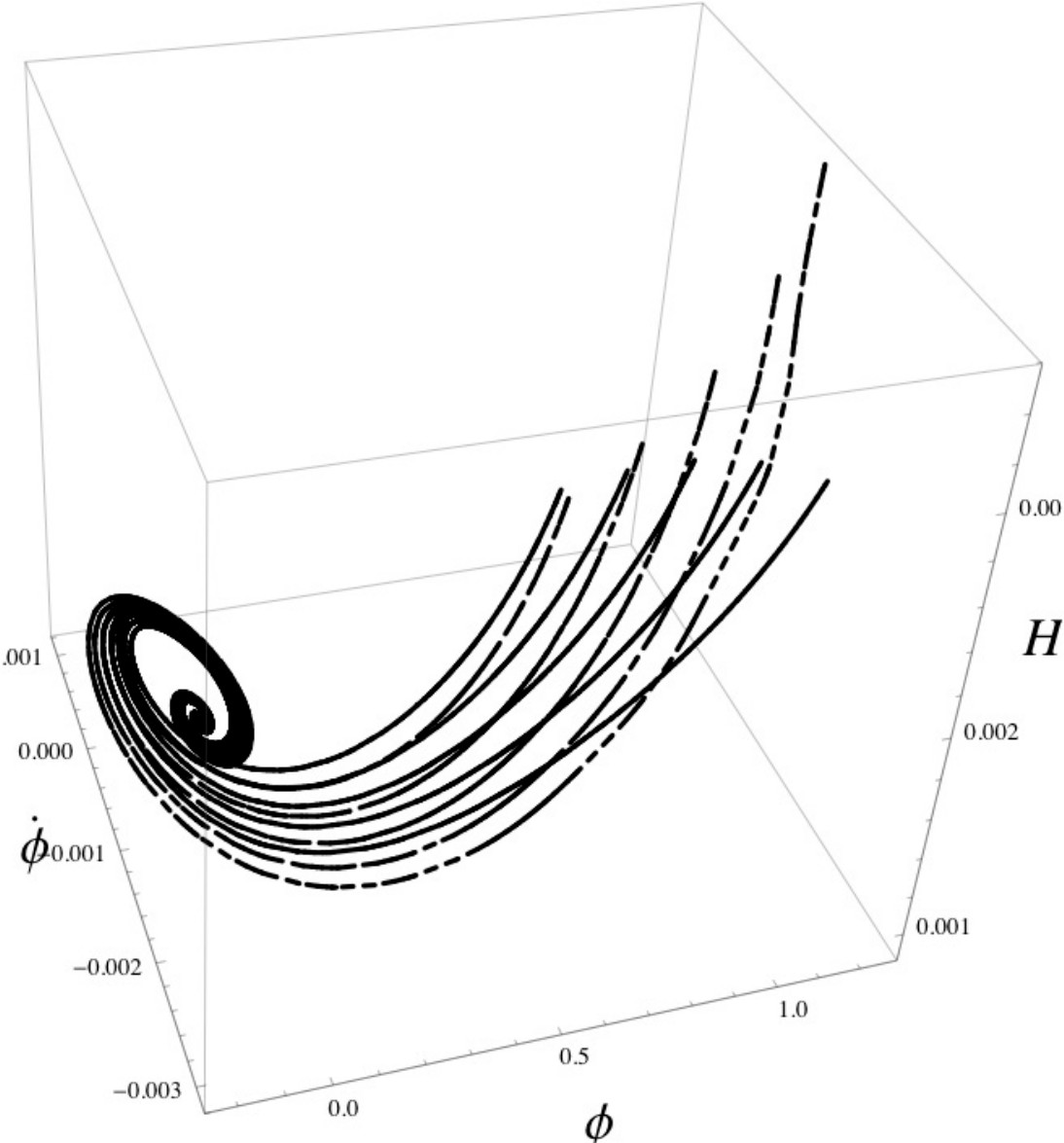}&
\includegraphics[scale=0.75]{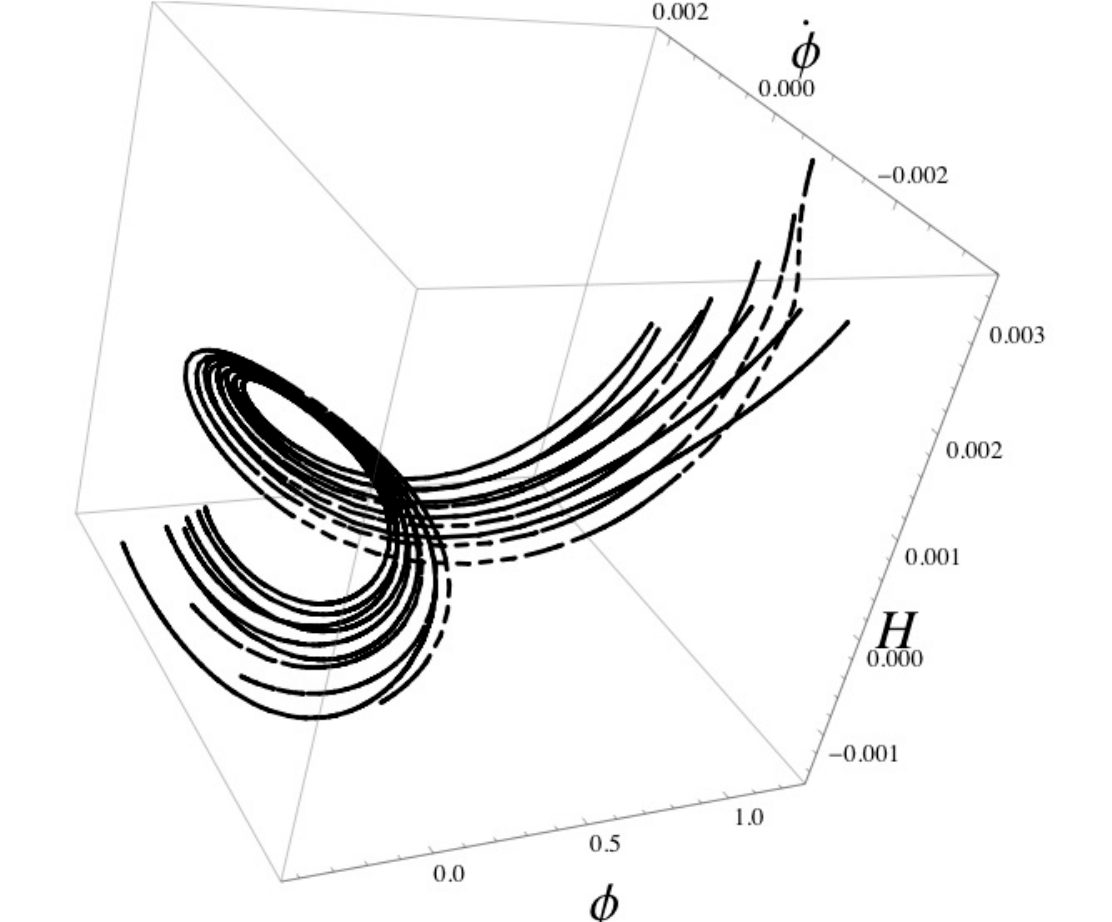}\\[0.0in] 
\end{array}$
\caption{Phase space topology in $\{\phi,\dot{\phi},H\}$. \textit{Left panel}: $\Lambda>0$, with evolution to a minimum $H$. \textit{Right Panel}: $\Lambda<0$, trajectories spiral through $H=0$ and the universe collapses. Dashed lines are for evolution in the full potential, and solid lines in the adiabatic potential, where the modulus remains always in its local minimum.}
\label{fig:3dphaseplots}
\end{figure*}
%%%%

We can obtain an effective potential for the axion that approximates the full effective field theory description by substituting the adiabatic trajectory for the modulus back into the potential of Eq.~(\ref{eqn:potential}):
\begin{align}
V_{\rm eff}(\phi)=&\frac{1}{2 \omega^2}(1-\frac{\beta^2}{6}\phi^2)  \bigg[  \frac{B}{2\omega^2}(1-\frac{\beta^2}{6}\phi^2) \nonumber \\
				&- D + \frac{M^2}{2}\phi^2 \bigg] + \rho_\Lambda\,.
\label{eqn:effective_potential}
\end{align}
This potential differs from the harmonic potential by becoming flat as $\phi \rightarrow \tilde{\phi}$. Beyond $\phi=\tilde{\phi}$ it should not be used.

During this evolution the fields follow a \emph{curved} trajectory in field space, just like in multi-field inflation, with the axion and modulus both always moving to smaller values and becoming heavier. Therefore the normal course of cosmic evolution will not endanger late time stability. However, as the bound of Eq.~(\ref{eqn:minbound}) becomes saturated we should see that axions and moduli undergo significant evolution in their masses while moving towards the global minimum. If the axion is light in the current epoch then this evolution will still be occurring.

Again drawing the analogy to multi-field inflation, even if the evolution in the modulus direction is slight, a tight turn in the field space trajectory may lead to observable features in the axion power spectrum~\cite{achucarro2011,cespedes2012}. Such tight turns do not appear possible in the potential we study, and we also will not be considering the effect of inhomogeneous perturbations.

We finally note here that if the amplitude of axion oscillations were allowed to grow, such as in the scenario explored in \cite{dubovsky2010} where axion oscillations are amplified by the Penrose process near a black hole, then this may also lead to novel effects on the modulus sector and the vacuum energy.

% -------------------------------------------------------------------------
\subsection{Dynamics in Phase Space and the Equation of State}

The value of the vacuum energy (which should not be confused with the scale of the potential in Section~(\ref{model})), $\Lambda = V(0,\chimin)$, which includes the $\rho_\Lambda$ contribution, is a free parameter in our model. Its sign controls the topology of phase space \cite{felder2002}. We show this effect in our model in Fig.~(\ref{fig:3dphaseplots}), where we plot trajectories in $\{ \phi,\dot{\phi}, H\}$ phase space obtained by numerically solving the equations of motion (Eqs.~(\ref{eqn:axionModuliEoM})). With $\Lambda > 0$ the trajectories are confined to the expanding branch, $H>0$ (or if $H_i<0$, the contracting branch), and the phase space is disconnected. With $\Lambda<0$ it is possible for the total energy density to go to zero, and so $H\rightarrow 0$ connecting the expanding and contracting branches and making phase space connected.

\begin{figure}
\centering
\includegraphics[scale=0.9]{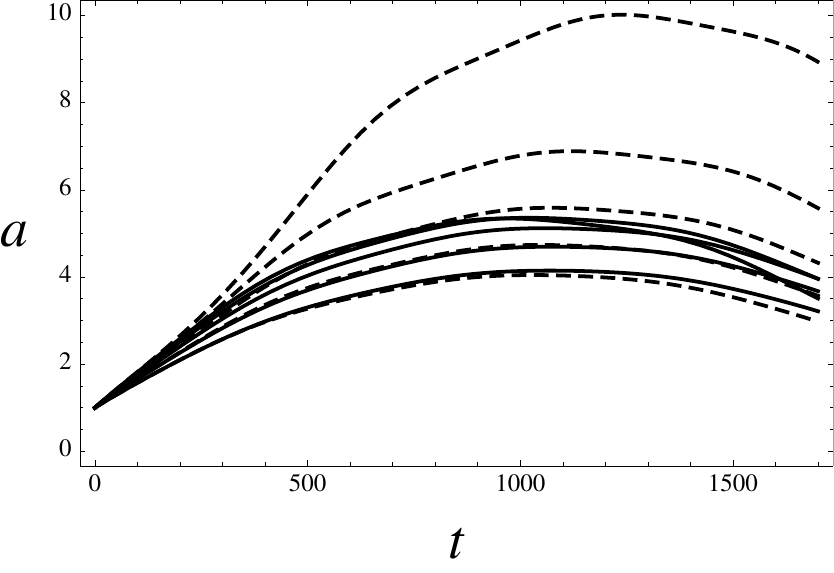}
\caption{The evolution of the scale factor for $\Lambda<0$, showing a turn over and collapse of the universe despite there being zero curvature. Again, dotted lines are for evolution in the full potential, while solid lines are in the adiabatic effective potential. Bottom to top corresponds to increasing $\phi_i \rightarrow \tilde{\phi}$. At large $\phi$ the universe reaches a larger size when the full potential is used.}
\label{fig:scale_plot_Ads}
\end{figure}

When $H=0$ the evolution of the scale factor turns over, such that with $\Lambda<0$ the universe expands and then contracts to a big crunch despite, in these cases, having zero curvature (see e.g. \cite{kallosh2003b} and references therein). We plot the evolution of the scale factor for $\Lambda<0$ in Fig.~(\ref{fig:scale_plot_Ads}).

The trajectories of Fig.~(\ref{fig:3dphaseplots}) all begin at the local modulus minimum, $\chimin (\phi_i)$, with stationary fields in a fluid dominated universe, and $\phi_i$ takes various values between $0$ and $\tilde{\phi}$. We begin at time $t=0, a=1$ during matter domination and look at the evolution towards the dark energy universe of today. We have shown trajectories given by evolution in the full potential (dashed lines), and in the effective potential (solid lines). Those trajectories with large initial axion field values in the full potential depart from the evolution in the effective potential. This is because at large axion values the modulus is light and the adiabatic assumption is no longer good enough. The scale factor evolution for $\Lambda<0$ is shown in Fig.~(\ref{fig:scale_plot_Ads}) and a clear difference is visible between evolution in the two potentials, with the maximum size of the universe being larger when the full potential is used.

\begin{figure}
\centering
\includegraphics[scale=1]{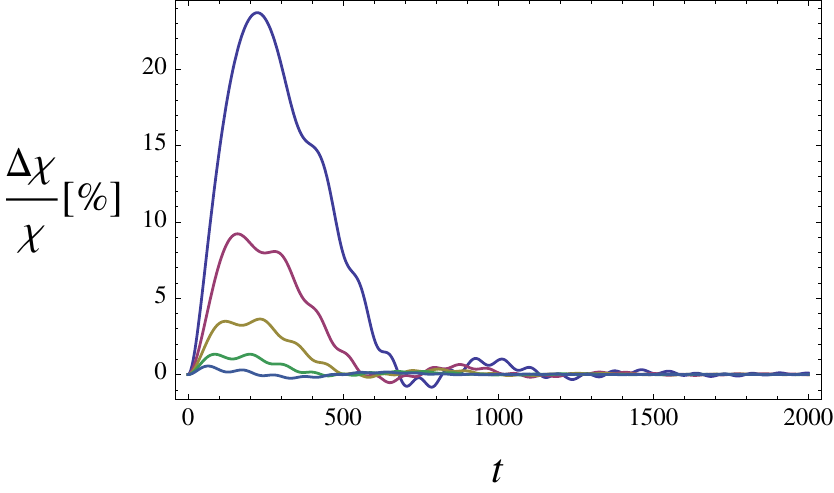}
\caption{Comparing modulus evolution for motion in the full potential versus motion in the adiabatic effective potential with $\Lambda>0$. The modulus always begins in the local minimum. Lines from bottom (light blue) to top (dark blue) represent increasingly large axion initial field values. The modulus departs by up to 25\% from the adiabatic trajectory when the initial value of the axion field is large.}
\label{fig:delta_chi_ds}
\end{figure}

We investigate the accuracy of the adiabatic approximation for $\Lambda>0$ in Fig.~(\ref{fig:delta_chi_ds}), where we give $\Delta \chi /\chi (t) = 1-\chimin(\phi)/\chi$ in percent. We see that at early times the trajectories with large initial $\phi$ depart by as much as 25\% from the adiabatic trajectory. All trajectories undergo damped oscillations about the adiabatic trajectory as the modulus mass increases over time.

%%%
%%%

%%%
%%%

\begin{figure}
\centering
\includegraphics[scale=1.0]{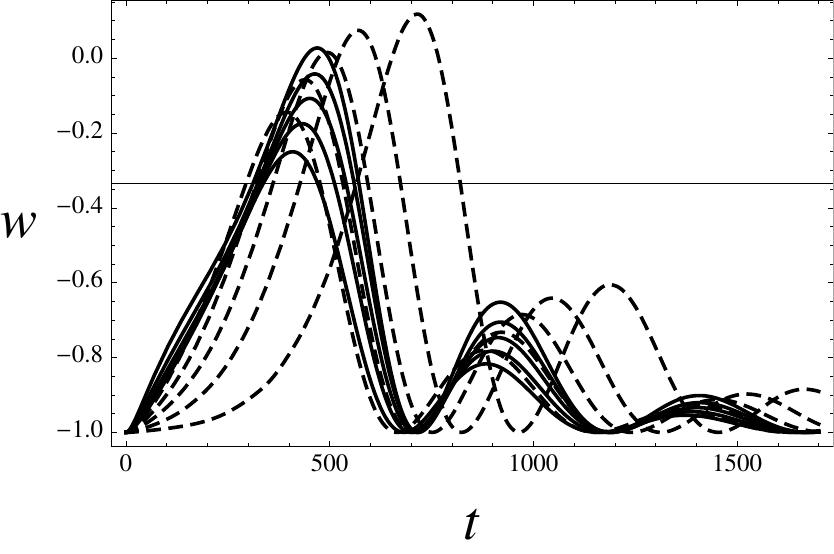}
\caption{The Dark Energy (DE) equation of state for $\Lambda>0$. Dotted lines are for motion in the full potential, solid lines for motion in the adiabatic effective potential. Left to right lines have increasing $\phi_i \rightarrow \tilde{\phi}$. DE has come to dominate the energy density around $t=100$, but later both the DE and total equation of state go back above $w=-1/3$ (horizontal line), temporarily halting accelerated expansion. In the most extreme case of large $\phi_i$ in the full potential this can happen even when the equation of state today is very flat and close to $w=-1$. The late time expansion is asymptotically dS as the fields relax into the minimum.}
\label{fig:eos_ds}
\end{figure}

We conclude this section by commenting on the effect of the combined axion-modulus oscillations about the minimum on the dark energy equation of state. With $\Lambda>0$ and at least one light field (conservatively, an axion of the axiverse), the fields at or close to their initial values can come to dominate the energy density as dark energy, yet still be evolving towards, and oscillating about, the true vacuum at late times. We show the effect of this on the equation of state $w_{\phi\chi\Lambda}=(\rho_\phi+\rho_\chi+\rho_\Lambda)/(P_\phi+P_\chi-\rho_\Lambda)$ in Fig.~(\ref{fig:eos_ds}), for the trajectories of Fig.~(\ref{fig:3dphaseplots}) (left panel), again comparing the cases of the effective potential and the full potential. 

In this evolution, the dark energy axion-modulus fluid is already dominating the energy density at early times, $t \sim 100$, with $w\approx -1$. However, $w$ is rising as the fields move towards the minimum leading to large departures from $w=-1$. For large axion initial values in the full potential this motion is delayed and $w$ remains flatter for longer. In all cases, as oscillations begin $w$ rises so much as to halt accelerated expansion altogether ($w>-1/3$). The positive value of $\Lambda$ in the true vacuum means that at late times $w$ will relax to exactly $-1$ and that $w$ is bounded to $-1\leq w \leq 1$. This bound does not hold for a negative potential \cite{wetterich2004}, indeed $|w_\chi|>1$ was observed for this potential in \cite{marsh2011}.

%%%
%%%

% -------------------------------------------------------------------------
\section{Dynamical Systems Analysis}\label{phasespace}

To delineate the regions of parameter space that may give rise to acceptable and interesting cosmological phenomenology we perform a dynamical systems analysis by transforming the coupled axion--modulus system into autonomous form. The axion and modulus fields and the perfect baryotropic fluid evolve according to Eqs.~(\ref{eqn:axionModuliEoM}), subject to the Friedmann constraint Eq.~(\ref{eqn:Friedmann}).

% -------------------------------------------------------------------------
\subsection{Autonomous System}

Following~\cite{copeland1998} we make the change of variables:
\begin{eqnarray}
	x \equiv \frac{\dot{\phi}}{\sqrt{6}H}\,,     &\quad&              y \equiv \frac{\dot{\chi}}{\sqrt{6}H}\,,   \nonumber \\
	z \equiv \frac{1}{H}\sqrt{{\frac{V_{B}}{3}}} \,,  &\quad& 	r  \equiv \frac{1}{H}\sqrt{{\frac{V_{D}}{3}}} \,,  \nonumber \\
	s  \equiv \frac{1}{H}\sqrt{{\frac{U}{3}}} \,,   &\quad&          t   \equiv \frac{1}{H}\sqrt{{\frac{\rho_{\Lambda}}{3}}} \,.
	\label{eqn:variables}	
\end{eqnarray}
The evolution Eqs.~(\ref{eqn:axionModuliEoM}) can then be transformed into autonomous form ${\bf X'}={\bf f(X)}$, where ${\bf X}$ is the column vector of compact variables and ${\bf f(X)}$ is the corresponding column vector constituting the autonomous system equations:
\begin{eqnarray}
	x' &=& -\left[\frac{H'}{H} +3 \right] x - \beta rs \,, \nonumber \\ 
         y' &=& -\left[\frac{H'}{H} +3 \right] y - \sqrt{\frac{3}{2}}C[r^2-s^2-2z^2] \,,  \nonumber \\
      	z' &=& -\left[\frac{H'}{H} +\sqrt{6}Cy \right] z \,,  \nonumber \\
	r' &=& -\left[\frac{H'}{H} +\sqrt{\frac{3}{2}}Cy \right] r \,,  \nonumber \\
	s' &=& -\left[\frac{H'}{H} +\sqrt{\frac{3}{2}}Cy \right] s + \beta xr \,,  \nonumber \\
	t' &=& -\frac{H'}{H}t \,, 
	\label{eqn:autonomousSystem}	
\end{eqnarray}
with
\begin{equation}
	\frac{H'}{H} = -\frac{3}{2}\gamma_{b}(1 - x^2 - s^2 - y^2 - z^2 + r^2 - t^2) -3x^2 -3y^2  \,,
	\label{eqn:constraint2}	
\end{equation}
where $\beta$ was defined in Eq.~(\ref{eqn:beta}).

Here, a prime denotes differentiation with respect to the number of $e$--foldings $N\equiv{\rm ln}\,(a)$. The dimensionless density parameters $\Omega_{i}\equiv \rho_{i}/3H^{2}$ of the cosmic components $i$ can be expressed as
\begin{equation}
	\Omega_{\phi}=x^{2}+s^{2}\,, \quad \Omega_{\chi}=y^{2}+z^{2}-r^{2}\,, \quad \Omega_{\Lambda}=t^{2} \,,
	\label{eqn:OmegaC}	
\end{equation}
and furthermore, flatness imposes 
\begin{equation}
	\Omega_{\rm b}=1-(x^2 + s^2 + y^2 + z^2 - r^2 + t^2) \,.
	\label{eqn:OmegaB}	
\end{equation}

\begin{center}
\begin{table*}[ht]
\hfill{}
\begin{tabular}{c|c|c|c|c|c|c|c}
\hline
\hline
	& $x_c$    &  $y_c$  &  $z_c$   &     $r_c$ &   $s_c$  &   $t_c$  &  Existence	\\
\hline
{\bf A}  & 0 	&  0	& 0	 & 	0 &    0  & 0  &  all $\beta,C,\gb$ \\
\hline
{\bf  B}   & 0 	&  0	& 0	 & 	0 &   0   & $\pm1$  & all $\beta,C,\gb$ \\
\hline
{\bf  C}   & $\pm\sqrt{1-y^2}$ 	&  $\pm y$	& 0	 & 	0 &   0   & 0  & $-1\leq y\leq1$ \\
\hline
{\bf  D}   & 0 	&  $\sqrt{\frac{2}{3}}C$  &  $\pm\sqrt{1-\frac{2}{3}C^2}$	 & 	0 &   0   & 0  &  $C\leq\sqrt{\frac{3}{2}}$ \\
\hline
{\bf  E}   & 0 	&  $\sqrt{\frac{3}{8}}\frac{\gamma_b}{C}$	& $\pm\frac{1}{4C}\sqrt{6\gamma_b(2-\gamma_b)}$ & 	0 &   0   & 0  & $\gb\leq2$  \\
\hline
{\bf  F}  & 0 	&  $\frac{C}{\sqrt{6}}$	& 0	 & 	0 &    $\pm\sqrt{1-\frac{C^2}{6}}$  & 0  & $C\leq\sqrt{6}$ \\
\hline
{\bf  G}  & 0 	&  $\sqrt{\frac{3}{2}}\frac{\gamma_b}{C}$	& 0	 & 0 &  $\pm\frac{1}{2C}\sqrt{6\gamma_b(2-\gamma_b)}$    & 0  & $\gb\leq2$ \\
\hline
%H ($\in\mathbb{C}$) & 0 	&  0	& $\pm i$ & 0 &  $\pm\sqrt{2}$    & 0  & -- \\
%\hline
{\bf  I}  & 0 	&  $\frac{C}{\sqrt{6}}$	& 0	 &  $\pm\sqrt{\frac{C^2}{6}-1}$ &    0  & 0  &  $C\geq\sqrt{6}$ \\
\hline
% J  & 0 	&  $\sqrt{\frac{3}{2}}\frac{\gamma_b}{C}$	& 0	 & $\pm\frac{1}{2C}\sqrt{6\gamma_b(\gamma_b-2)}$ &    0  & 0  &  $\in\mathbb{C}$ for realistic $\gb$ \\
%\hline
% K ($\in\mathbb{C}$)  & 0 	&  0	& $z$	 & 0 &    $\pm i\sqrt{2}z$  & $\pm\sqrt{z^2+1}$  &  -- \\
%\hline
% L ($\in\mathbb{C}$)  & 0 	&  0	& $\pm i$	 & $\pm i\sqrt{2}$ &    0  & 0  &  -- \\
%\hline
{\bf  M}   & 0 	&  0	& $\pm z$  & $\pm\sqrt{2}z$ &   0   & $\pm\sqrt{z^2+1}$  & all $\beta,C,\gb$ \\
\hline
\end{tabular}
\hfill{}
\caption{The fixed points of the system~(\ref{eqn:autonomousSystem}) and the conditions for their existence. Rather than having an isolated fixed point, ${\bf M}$ is formed of a continuous line of fixed points, called a critical line. This critical line intersects the $\zeta$ plane at a unique point $z=z_M$ given by Eq.~(\ref{eqn:zetaCrossing}).}
\label{tab:fixedPoints}
\end{table*}
\end{center}

\begin{figure}
\centering
\includegraphics[scale=0.8]{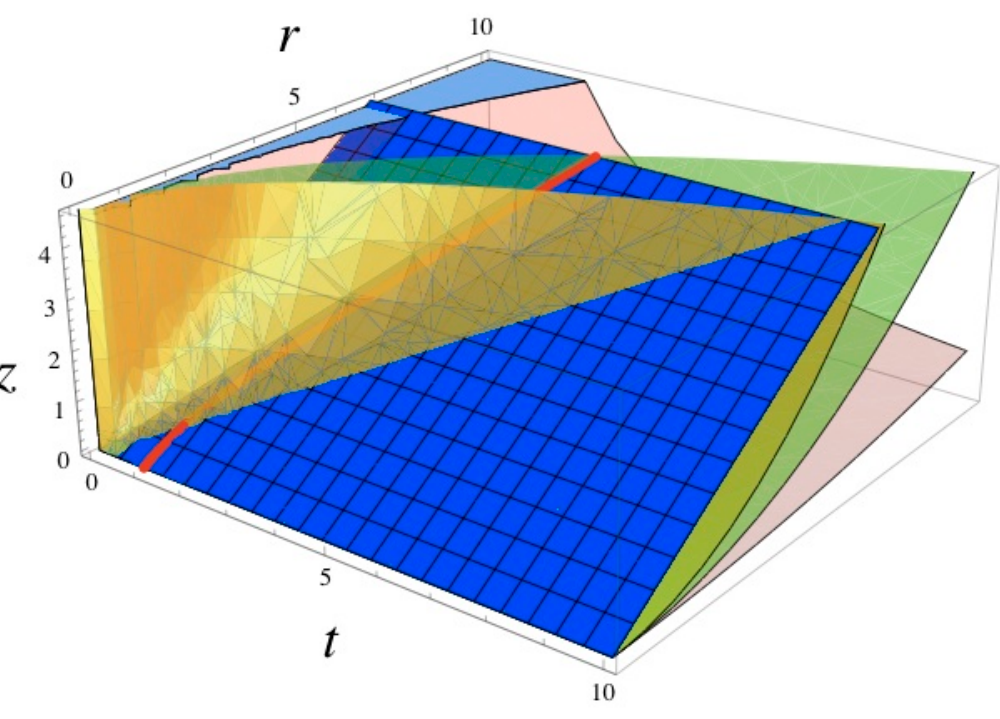}
\caption{Dynamical constraint surfaces in the $\{z,r,t \}$ subspace. The flat meshed (blue), surface corresponds to the ${\bf \mathcal{Z}}$ plane, which is the minimum of the potential. The foremost curved (yellow) surface is the constraint for a dS vacuum with $\zeta>1/4$. Next behind it is the Minkowski plane (green) with $\zeta=1/4$. Finally, the rearmost curved surface (red) is the constraint for an AdS vacuum with $\zeta<1/4$. The bold (red) line on the ${\bf \mathcal{Z}}$ surface corresponds to the critical line ${\bf M}$. We see that ${\bf M}$ crosses only dS planes and asymptotes to the Minkowski plane at co--ordinate infinity, while the minimum surface crosses all $\zeta$ planes on a line. The non-crossing of fixed line ${\bf M}$ with surfaces of $\zeta<1/4$ is another geometric illustration of the instability of a negative potential to collapse.}
\label{fig:dynamical_surfaces}
\end{figure}

At this point some comments on the system~(\ref{eqn:autonomousSystem}) are in order. Notice that due to the negative contribution from $V_D$ in the modulus potential, trajectories are not confined to the unit hypersphere in the full phase space. The set of phase space variables $\{ x,y,z,r,s,t \}$ is of one dimension more than the actual $\{ \phi,\dot{\phi},\chi,\dot{\chi},H \}$ degrees of freedom. This is because there is a relation that exists between the phase space variables, which provides an additional constraint and defines a surface on which the motion takes place, just like the Friedmann constraint gives the topology of phase space in~\cite{felder2002}. The constraint is:
\begin{equation}
	\frac{t^2 z^2}{r^4}=\frac{B\rho_\Lambda}{D^2}\equiv \zeta\,,
\label{eqn:defZeta}
\end{equation}
which is a simple consequence of the definitions of the variables and the form of the potential. Trajectories are confined to live on this plane, defined by the choice of initial conditions. There is a simple interpretation of this that will help us visualise phase space: Choosing $\zeta$ corresponds to a choice of sign for the vacuum energy, including the bare cosmological constant. Combining Eqs.~(\ref{eqn:potential}) and (\ref{eqn:chimin}) and $\phi=0$ to get the vacuum energy $\langle V \rangle = \Lambda$:
\begin{equation}
\Lambda = \frac{D^2}{B} \left( \zeta- \frac{1}{4} \right) \, .
\label{eqn:vacuum_cc}
\end{equation}
The Minkowski vacuum is given by $\zeta=\frac{1}{4}$, whilst $\zeta<\frac{1}{4}$ is an AdS vacuum, and $\zeta>\frac{1}{4}$ is a dS vacuum (we show the planes for three values of $\zeta$ in Fig.~(\ref{fig:dynamical_surfaces}), which we will discuss in more detail later), but we note that $\zeta$ does not set the scale of the vacuum energy. This would be fixed observationally \emph{if} we were to want the vacuum to give us the correct $H_0$, but this is not necessary: $\langle V \rangle$ can be much less than this, but not greater. 

Furthermore, this geometrical picture of surfaces in the $\{ z,r,t \}$ subspace can give another view on the boundedness of trajectories and topology of phase space discussed in Section (\ref{potential}) in Fig.~(\ref{fig:3dphaseplots}). The requirement of a flat universe imposes the constraint:
\begin{equation}
x^2+y^2+s^2+z^2-r^2+t^2 \leq 1\,.
\end{equation}
Clearly, the location of the hypersurface defined by saturation of the bound moves as the variables evolve, however we can picture its effect in the limit of heading to the vacuum: $x=y=s=0$. Now there is an additional surface that intersects those of Fig.~(\ref{fig:dynamical_surfaces}). If one were to plot it, one would see that it intersects dS surfaces, with $\zeta>1/4$, making an arc below which trajectories are confined, unable to reach co--ordinate infinity. For AdS surfaces, with $\zeta<1/4$, the surface funnels outwards, restricting trajectories to a region of their $\zeta$ surface, but not confining them to finite values. This is another manifestation of our choice of dynamical system variables: for a negative vacuum energy it is possible for $H \rightarrow 0$, where the co-ordinates diverge and trajectories on the potential become unconfined.

%%%
%%%

The plane defined by $\zeta=\frac{1}{4}$, which we will call the Minkowski plane, therefore divides the phase space into three: above the Minkowski plane, phase space is bounded, and the collapsing and expanding universe branches ($z,r,s,t<0$ and $z,r,s,t>0$ respectively) are separated; on the Minkowski plane the asymptotic future lies at infinity, where $H=0$; below the Minkowski plane, phase space is connected and there are trajectories through infinity that join the expanding and contracting branches. When the transition is made from expansion to contraction and $H$ changes sign, all six variables, $x,y,z,r,s,t$ also change sign. 

Choosing to work with a phase space of one higher dimension is useful for our analysis of fixed points, since it allows us to see what happens when one or more of these variables can be approximated as vanishing relative to the others, for example $z$ and $r$ vanish as the modulus goes to large values, and $t$ vanishes in the early universe. 
\begin{center}
\begin{table*}[htp]
\hfill{}
\begin{tabular}{c|c|c|c|c|c|c|c}
\hline
\hline
	& $\Omega_\phi$    &  $\Omega_\chi$  &  $\Omega_\Lambda$   &     $\Omega_b$ & $\ddot{a}>0$?  &  $w_{\rm eff}$   & 	Stability\\
\hline
{\bf A}      &    0  & 0 & 0 & 1  & never & $0$  &  unstable \\
\hline
{\bf B}      &    0  & 0 & 1 & 0   & always &  $-1$ &  marginally stable  \\
\hline
{\bf C}      &    $1-y^2$  & $y^2$ & 0 & 0 & never  & $1$  &  unstable \\
\hline
{\bf D}      &    0  & 1 & 0 & 0  & $C<\sqrt{\frac{1}{2}}$ &  $-1+\frac{4}{3}C^2$  &  unstable\\
\hline
{\bf E}     &    0  & $\frac{3}{4}\frac{\gamma_b}{C^2}$ & 0 &  $1-\frac{3}{4}\frac{\gamma_b}{C^2}$ & never  & $\frac{3}{4}\frac{\gb(\gb-1)}{C^2}$  &  unstable \\
\hline
{\bf F}      &    $1-\frac{C^2}{6}$  & $\frac{C^2}{6}$ & 0 & 0 & $C<\sqrt{2}$  &  $-1+\frac{1}{3}C^2$   &  unstable \\
\hline
{\bf G}      &    $\frac{3}{2}\frac{\gamma_b(2-\gamma_b)}{C^2}$  & $\frac{3}{2}\frac{\gamma_b^2}{C^2}$ & 0 &  $1-\frac{3\gamma_b}{C^2}$ & never  &  $\frac{3\gb(\gb-1)}{C^2}$ &  unstable   \\
\hline
%H ($\in\mathbb{C}$)      &    2  & -1 & 0 &  0  & --  & --  &  -- \\
%\hline
{\bf I}      &    0  & 1 & 0 & 0 & $C<\sqrt{2}$  & $-1+\frac{1}{3}C^2$ &  unstable\\
\hline
%J      &    0  & $\frac{3\gamma_b}{C^2}$ & 0 & $1-\frac{3\gamma_b}{C^2}$ &  & $\frac{3\gb(\gb-1)}{C^2}$  &  unstable\\
%\hline
%K ($\in\mathbb{C}$)      &    $-2z^2$  & $z^2$ & $z^2+1$ & 0 & -- & --   & --  \\
%\hline
%L ($\in\mathbb{C}$)      &    0  & 1 & 0 & 0 & -- &  --  & --\\
%\hline
{\bf M}      &    0  & $-z^2$ & $z^2+1$ & 0 & always  & $-1$ &  stable\\
\hline
\end{tabular}
\hfill{}
\caption{Properties of the fixed points given in Table~(\ref{tab:fixedPoints}) for an expanding universe. For ${\bf M}$, the contribution from $\rho_\Lambda$ has been included in $w_{\rm eff}$.}
\label{tab:fixedPointDensities}
\end{table*}
\end{center}
When $H=0$ our variables diverge and so the system~(\ref{eqn:autonomousSystem}) cannot be evolved through the transition between expanding ($H^+$) and contracting ($H^-$) universes. It is actually possible to construct a set of compact variables which remain finite at $H=0$: 
\begin{eqnarray}
	x_Q \equiv \frac{\dot{\phi}}{\sqrt{2}Q}\,,     &\quad&              y_Q \equiv \frac{\dot{\chi}}{\sqrt{2}Q}\,,   \nonumber \\
	z_Q \equiv \frac{\sqrt{V_B}}{Q} \,,  &\quad& 	s_Q \equiv \frac{\sqrt{U}}{Q} \,,  \nonumber \\
	t_Q  \equiv \frac{\sqrt{\rho_{\Lambda}}}{Q} \,,      
	\label{eqn:Qvariables}	
\end{eqnarray}
where 
\begin{equation}
	Q \equiv \sqrt{3H^2 + V_D}\,.
	\label{eqn:Qdef}	
\end{equation}
These variables are similar to those defined in~\cite{Coley2000Closed}. Since $V_D$ is positive definite, $Q$ always remains well defined. Defining a new independent variable $(')=\frac{1}{Q}\frac{\rm d}{{\rm d}t}$ one can transform the evolution Eqns.~(\ref{eqn:axionModuliEoM}) and~(\ref{eqn:Friedmann}) into autonomous form. This alternative autonomous system is given in Appendix~(\ref{appdx:autonomousSystemQ}). These compact variables, ${\bf X_Q}$, are related to our original compact variables ${\bf X}$ (Eqs.~(\ref{eqn:variables})) by
\begin{equation}
	{\bf X}={\bf X_Q} \hat{r}\,, \quad \quad \hat{r}=\sqrt{1+r^2}\,,
	\label{eqn:XtoXQ}	
\end{equation}
where $r$ was defined in Eqs.~(\ref{eqn:variables}). We have kept our variables finite at $H=0$, at the expense of losing the intuitive description of the division of phase space provided by the vacuum constraint Eq.~(\ref{eqn:defZeta}), since the ${\bf X_Q}$ system has the minimally required dimensionality. For this reason we content ourselves with describing the axion--modulus system in terms of the variables of Eqs.~(\ref{eqn:variables}), and do not study the transition at $H=0$ explicitly. We will occasionally make use of the ${\bf X_Q}$ variables to numerically show the evolution of phase space trajectories.\\

% -------------------------------------------------------------------------
\subsection{Fixed Points}\label{fixedpoints}

The fixed (critical) points ${\bf X_c}$ of the autonomous system~(\ref{eqn:autonomousSystem}) are extracted by satisfying ${\bf X'=0}$ and are listed, along with their conditions for existence, in Table~(\ref{tab:fixedPoints}). As mentioned earlier, the positive (negative) roots in the $\{z,r,s,t\}$ subspace correspond to expanding (contracting) universes. In total there are thirteen fixed points, four of which are imaginary and so are not physical and are not listed in Table~(\ref{tab:fixedPoints}). The energy densities $\Omega_i$, the effective scalar field equation of state, $w_{\rm eff}=(P_\phi+P_\chi)/(\rho_\phi+\rho_\chi)$ and conditions for acceleration and stability of these fixed points are given in are given in Table~(\ref{tab:fixedPointDensities}). A fixed point corresponds to an accelerating solution if
\begin{equation}
	(1-x_c^2-y_c^2-s_c^2-z_c^2+r_c^2-t_c^2)\gb+2x_c^2+2y_c^2 < \frac{2}{3}\,.
	\label{eqn:FPaccel}	
\end{equation}

The stability of the fixed points may be determined by expanding about them, setting ${\bf X=X_c+\delta X}$, with ${\bf \delta X}$ the perturbations of the compact variables defined by Eqs.~(\ref{eqn:variables}) considered as a column vector. To first order, the perturbations satisfy ${\bf \delta X'=W\cdot \delta X}$, where the matrix ${\bf W}$ contains the coefficients of the perturbation equations. The stability of the fixed points thus depends upon the nature of the eigenvalues of the matrix ${\bf W}$. The full stability analysis is somewhat cumbersome and may be found in Appendix~(\ref{appdx:stability}). Here, we give a general summary of the fixed points and their stability, focussing on the intuitive physics that the dynamical systems approach provides.

\begin{figure*}[t]
	\begin{tabular}{cc}
		\includegraphics[width=8.5cm]{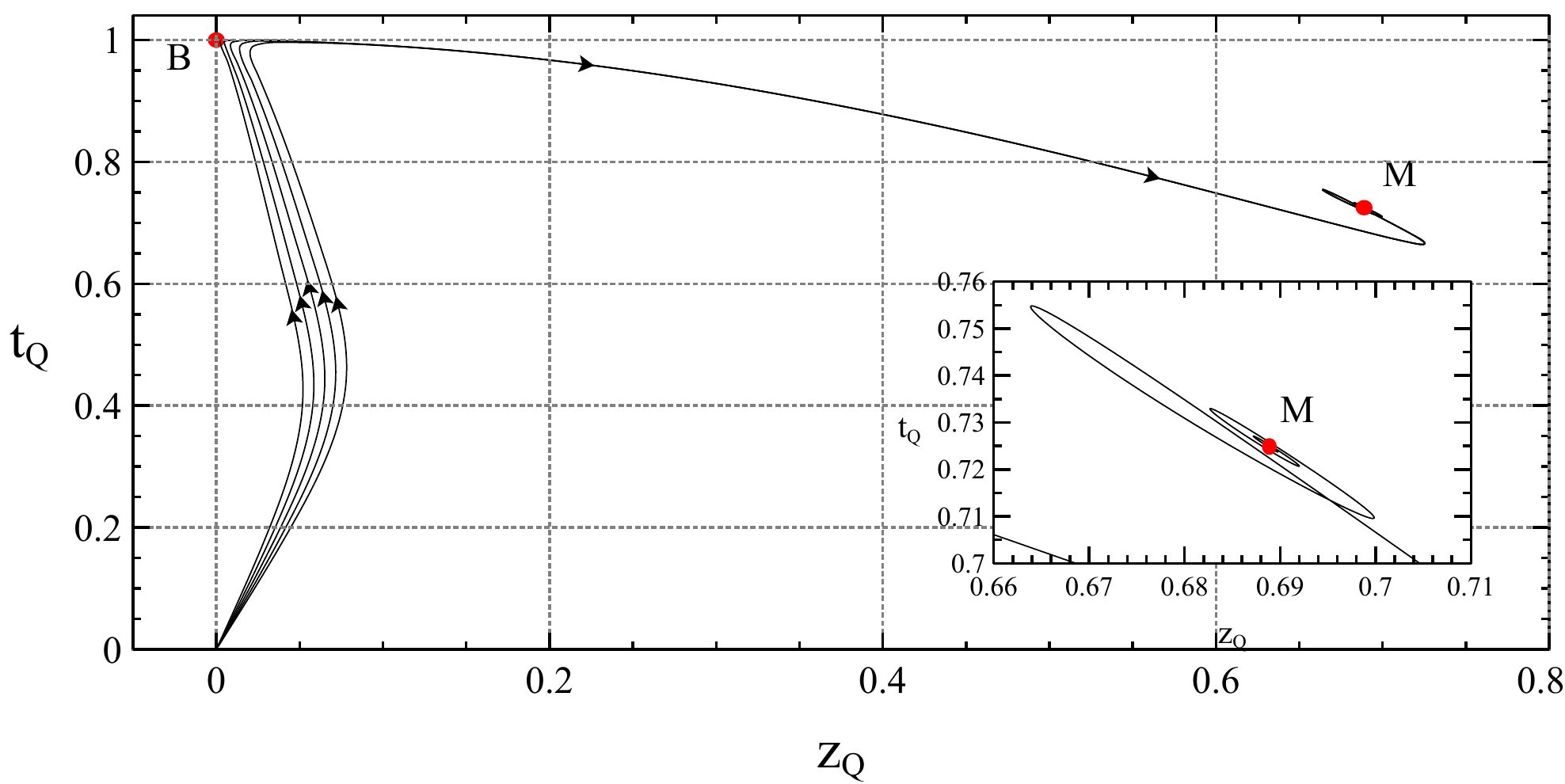} &
		\includegraphics[width=8.5cm]{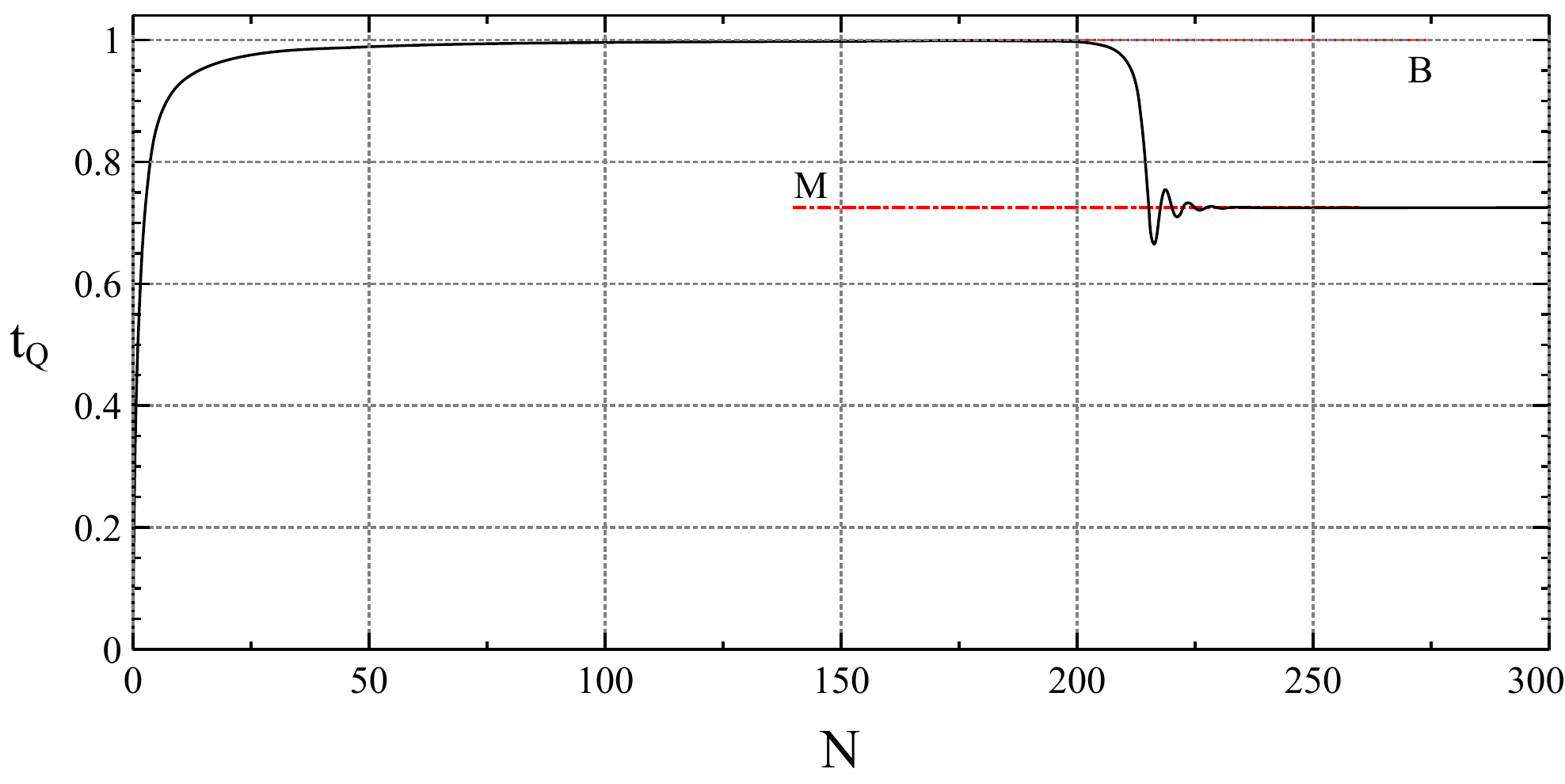} \\
	\end{tabular}
	\caption{The evolution of phase space trajectories in the presence of a dS vacuum ($\zeta=0.276$) obtained by integrating Eqs.~(\ref{autonomousQ}). We set $C=1$, $\gb=1$ and $\beta=1.87$. The compact variables $z_Q$ and $t_Q$ are related to $z$ and $t$ through Eq.~(\ref{eqn:XtoXQ}). \textit{Left panel:}  The temporary trapping of the $z_Q$ and $t_Q$ trajectories in fixed point ${\bf B}$ before the modulus begins to roll, finding its minimum at fixed point ${\bf M}$. Saturation of the bound $C\leq\sqrt{\frac{3}{2}\left(\zeta-\frac{1}{4}\right)}$, Eqn.~(\ref{eqn:bifurcationMatCrossing}), results in late--time modulus oscillations, which are seen in the figure as trajectories spiralling into ${\bf M}$.  \textit{Right panel:} $t_Q$ as a function of $N$, the number of $e$--foldings.}	
	\label{fig:phasePortraitsQ}
\end{figure*}

Of the nine fixed points listed in Table~(\ref{tab:fixedPoints}), there are two trivial solutions: Fixed point ${\bf A}$ corresponds to the fluid dominated point where the kinetic and potential components of the axion and modulus fields are negligible, whilst fixed point ${\bf B}$ represents the $\rho_\Lambda$ dominated solution. Point ${\bf A}$ is unstable in both expanding and contracting universes. Recall the ultimate fate of the universe is determined by the value of $\zeta$. In the presence of a dS vacuum ($\zeta>1/4$) the stability analysis reveals that fixed point ${\bf B}$ is associated with three \textit{zero} eigenvalues in the $\{z,r,s\}$ subspace, whilst the remaining directions are stable. We say that this is a \textit{marginally stable} solution in the sense that there is no instability growing exponentially, although it could be unstable to higher orders in the perturbation. To obtain the strict stability of this solution we would have to go beyond linear order in perturbation theory, which we do not pursue as numerical integration of the autonomous system confirms that this point is ultimately unstable: the asymptotic future  in the presence of a dS vacuum is the stable fixed point ${\bf M}$, the global axion--modulus potential minimum, which has a larger basin of attraction. The existence of point ${\bf B}$ demonstrates the ability of a bare cosmological constant to overdamp modulus motion for the modulus beginning life high up on the plateau of its potential, shielding us from the true vacuum and seeing only the larger $\rho_\Lambda$. In Fig.~(\ref{fig:phasePortraitsQ}) we show this temporary `trapping' in fixed point ${\bf B}$ by plotting trajectories in the $\{z_Q,t_Q\}$ subspace. Whilst such a trapping may last for hundreds or even thousands of $e$--foldings, the modulus will eventually begin to roll when its mass overcomes the Hubble damping and will relax into its minimum.

\begin{figure*}[t]
	\begin{tabular}{cc}
		\includegraphics[width=8.5cm]{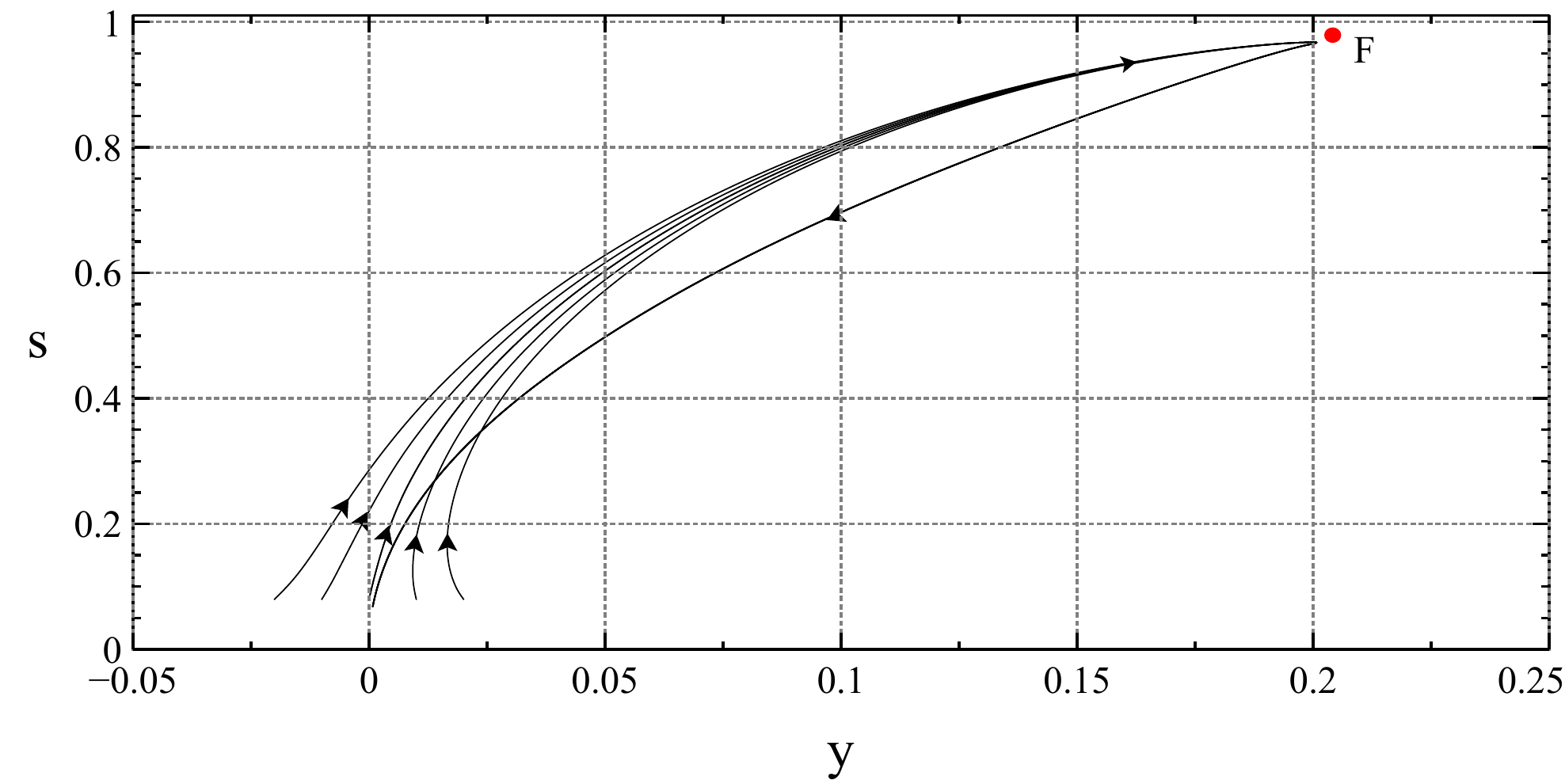} &
		\includegraphics[width=8.5cm]{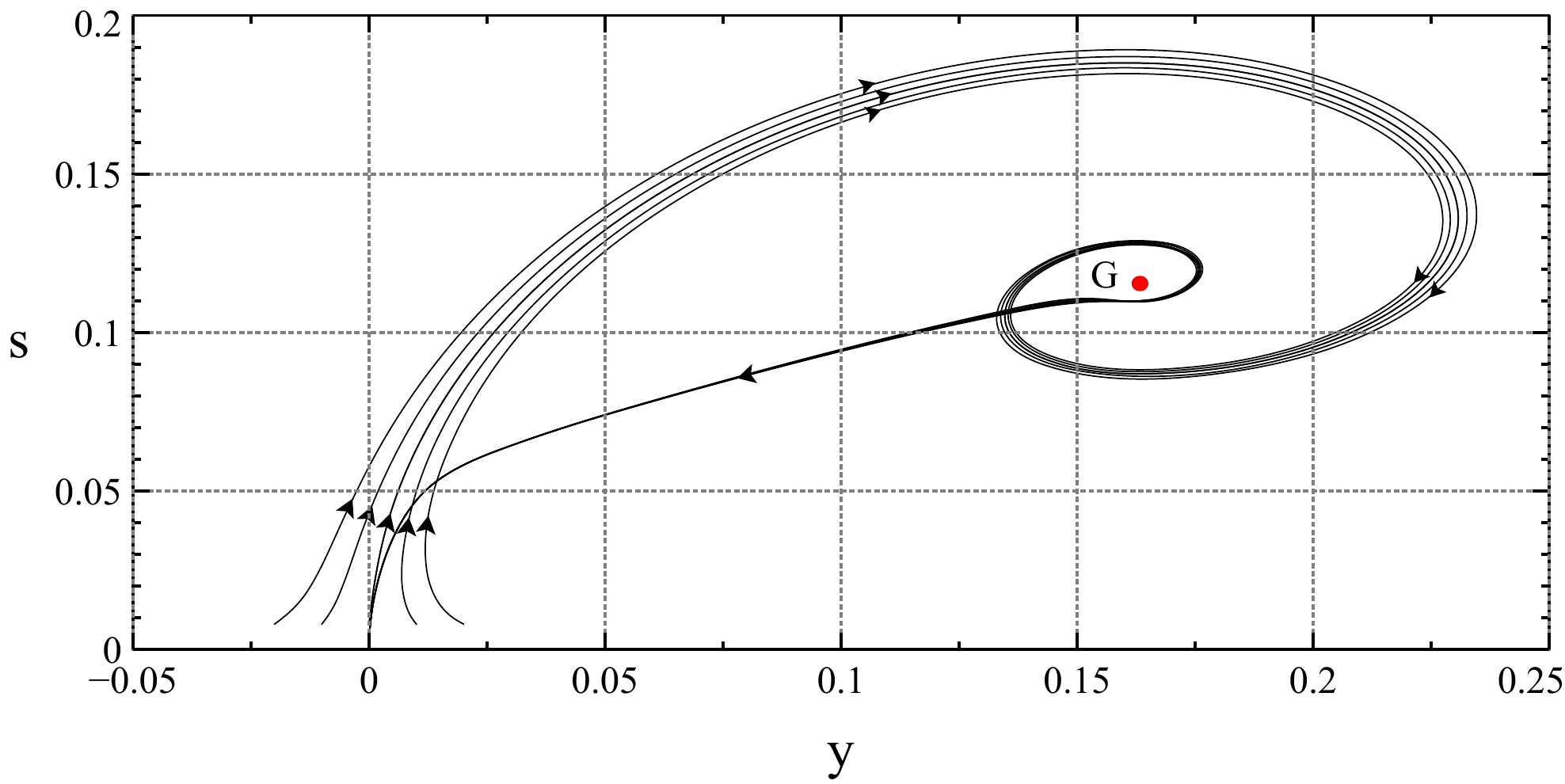} \\
	\end{tabular}
	\caption{The evolution of phase space trajectories in the presence of a dS vacuum obtained by integrating Eqs.~(\ref{eqn:autonomousSystem}). \textit{Left Panel}: Trajectories repelled from the unstable fixed point ${\bf F}$ and heading for the global potential minimum at ${\bf M}$. We set $\zeta=1.0$, $C=0.5$, $\gb=4/3$ and $\beta=1.732$. \textit{Right Panel}: Trajectories spiralling into the unstable fixed point ${\bf G}$, before the modulus finds the global potential minimum at ${\bf M}$. We set $\zeta=1.0$, $C=10.0$, $\gb=4/3$ and $\beta=10.0$.}	
	\label{fig:phasePortraitsQ2}
\end{figure*}

Rather than having an isolated fixed point, point ${\bf M}$ is formed of a continuous line of fixed points, known as an \textit{equilibrium manifold}, which we call a \textit{critical line}. The emergence of this critical line is due to the fact that we are working in one dimension more than is required. In one dimension less, the line would degenerate to a unique point, which is given by the intersection of ${\bf M}$ with the $\zeta$--plane,
\begin{equation}
	z_M = \pm \frac{1}{\sqrt{4 \zeta -1}}\,,
	\label{eqn:zetaCrossing}
\end{equation}
which clearly only exists for $\zeta>\frac{1}{4}$. That is to say, the absolute potential minimum defined by the stable fixed point ${\bf M}$ is only a fixed point in the presence of a dS vacuum and corresponds to the asymptotic future. If $\zeta<\frac{1}{4}$ the global minimum is not a fixed point, and the asymptotic future is cosmic doomsday in a Big Crunch as will be discussed below. The global minimum is expressed in terms of the autonomous system variables by substituting $z=z_M$ in ${\bf M}$. We finally note that the line ${\bf M}$ can equally be derived as the intersection of the minimum surface with the saturation of the flatness constraint in the potential dominated regime, $z^2-r^2+t^2=1$.

As well as confirming that ${\bf M}$ is stable for $\zeta>\frac{1}{4}$, the stability analysis reveals two \textit{bifurcation points} that lie along it:
\begin{equation}
 	z=\sqrt{\frac{3}{8C^2}}\,, \quad {\rm and} \quad z=\sqrt{\frac{9}{8\beta^2}}\,.
	\label{eqn:bifurcationM}
\end{equation} 
These bifurcation points are obtained by setting the quantity under the square root in the eigenvalues $\lambda_{1,2}$ and $\lambda_{4,5}$ of Eq.~(\ref{EigenM}) to zero and solving for $z$. For {\bf M} to be a stable node in the $\{y,z,r\}$ subspace, $z\leq\sqrt{3/8C^2}$, otherwise it is a stable spiral, whilst for point {\bf M} to be a stable node in the $\{x,s\}$ subspace, $z\leq\sqrt{9/8\beta^2}$, otherwise it is a stable spiral. The vacuum surface $\zeta$, determined by $B$, $\rho_\Lambda$ and $D$, dictates which side of the bifurcation points $z_M$ lies. We have the conditions 
\begin{equation}
 	\beta\leq\sqrt{\frac92\left(\zeta-\frac{1}{4}\right)}\,, \quad {\rm and} \quad C\leq\sqrt{\frac32\left(\zeta-\frac{1}{4}\right)}\,,
	\label{eqn:bifurcationMatCrossing}
\end{equation}
which are derived by setting $z=z_M$ in Eq.~(\ref{eqn:bifurcationM}). The stability analysis has elegantly revealed the conditions for late--time oscillations of the axion and modulus fields: violation of the $\beta$ condition corresponds to axion oscillations, whilst violation of the $C$ condition corresponds to modulus oscillations. Fig.~(\ref{fig:phasePortraitsQ}) shows an example of late--time modulus oscillations as the trajectories spiral into the point $z_M$. The time scale of the axion oscillations are determined by $\beta$, $C$ and also $\phi_i$. If the vacuum is chosen to be Minkowski, the conditions~(\ref{eqn:bifurcationMatCrossing}) simply become $\beta\leq0$ and $C\leq0$, which are never satisfied for the parameter values considered in this work and so axion and modulus oscillations are inevitable. 

There is another plane which is of interest in the $\{z,r,t\}$ subspace, which defines the minimum of the potential itself at $\chimin$. At $\chimin$, $r=\sqrt{2}z$, which defines the plane, and the critical line ${\bf M}$ lives here. This plane, which we will call ${\bf \mathcal{Z}}$, crosses the $\zeta$ plane on a line. Trajectories along this line are those living in the minimum and leading to the asymptotic future, either at the crossing point of ${\bf M}$ in a dS vacuum, or ultimately leading to collapse in an AdS vacuum. Trajectories crossing this line are modulus passages through, or oscillations about, the minimum. Trajectories in the full 6--d space, however, never cross each other: these are oscillations and static passages along the ${\bf \mathcal{Z}}$--$\zeta$ crossing and are separated in the $y$--direction. These surfaces are shown in the expanding octant of the $\{z,r,t\}$ plot in Fig.~(\ref{fig:dynamical_surfaces}), where we see the crossing of ${\bf \mathcal{Z}}$ along a line in the dS, Minkowski and AdS example $\zeta$ planes, ${\bf M}$ lying in the ${\bf \mathcal{Z}}$ plane, and crossing the dS plane at a point.

\begin{figure}
\centering
\includegraphics[scale=0.29]{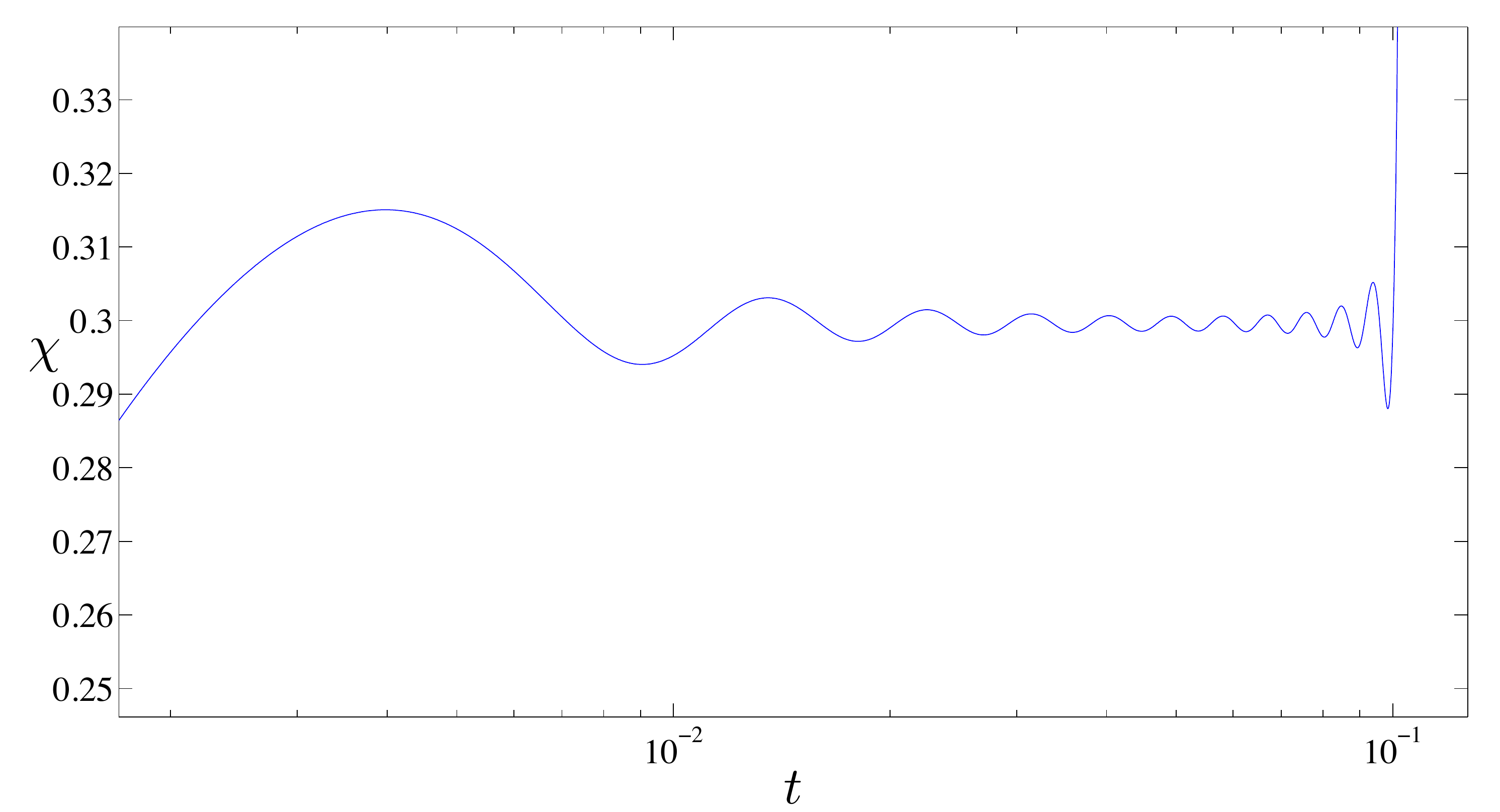}
\caption{Evolution of the modulus field in a collapsing universe. As $a\rightarrow 0$ at $t\sim 10^{-1}$ the kinetic energy grows and dominates in fixed point ${\bf C}$ and the field value diverges. Since this would happen to \emph{all} scalar fields and hence all moduli, this signals decompactification near a crunch. In this example plot the units and parameter values are all arbitrary.}
\label{fig:mod_field_collapse}
\end{figure}

Fixed point ${\bf C}$ is the second critical line of the system, corresponding to an axion--modulus kinetic dominated (stiff fluid) solution. This critical line is the unit circle $x_c^2+y_c^2=1$ and is a symmetry of the autonomous system with $z_{\rm c}=r_{\rm c}=s_{\rm c}=t_{\rm c}=0$. This is the usual enhancement of symmetry for massless scalar fields. In an expanding universe (where stability is ensured by negative eigenvalues of ${\bf W}$), this point is always unstable. For a collapsing universe, a fixed point is stable if the eigenvalues of ${\bf W}$ are \textit{positive}. This is because the `time' variable $N\equiv{\rm ln}\,(a)$ of the autonomous system becomes a decreasing function of time. Hence, critical line ${\bf C}$ is stable in a collapsing universe and corresponds to the asymptotic future of any model with an AdS vacuum. This is consistent with the pre--big bang cosmology~\cite{gasperini1993,lidsey1999} late time attractor solutions. The particular fixed point along {\bf C} that the system will finally evolve to will depend upon the initial conditions of the system. Similarly to the phase space dynamics in the presence of a dS vacuum, the only possibility to save us from this Big Crunch cosmic doomsday is a temporary trapping in fixed point ${\bf B}$. This situation was seen in the examples of~\cite{marsh2011} whenever $\rho_\Lambda$ domination sets in before collapse and is achieved for large initial modulus values. This period of dS inflation would only be temporary however and the modulus will quickly relax into its AdS vacuum signalling cosmological collapse and leading to eventual decompactification as described in~\cite{giddings2004}. This is demonstrated in the example plot of Fig.~(\ref{fig:mod_field_collapse}): the growing kinetic energy of the modulus as $a\rightarrow 0$ drives it to large values.

All other fixed points are unstable in the presence of an AdS or dS vacuum. Points ${\bf D}$ and ${\bf E}$ correspond to dynamical modulus stabilisation at small modulus values, while ${\bf I}$ corresponds to dynamical stabilisation at large field values. This is of course only a meta--stability, since these fixed points are unstable. Point ${\bf E}$ is also a scaling solution, on which the axion energy density vanishes ($\rho_\phi=0$) and the modulus energy density scales with the dominant background fluid:
\begin{equation}
	\rho_\chi=\frac{9H_i^2}{4C^2}\gb\left(\frac{a}{a_i}\right)^{-3\gb}\,, \quad w_\chi=\frac{3}{4}\frac{\gb(\gb-1)}{C^2}\,.
	\label{eqn:FPEscalingSol}
\end{equation}
Hence, the modulus tracks the dominant background fluid and $\rho_\chi/\rho_{\rm b}$ remains constant. 

Fixed point ${\bf F}$ represents a solution dominated by the modulus kinetic energy and the potential energy of the axion. On this solution, their relative energy densities and effective equation of state remains fixed:
\begin{equation}
	\frac{\rho_\phi}{\rho_\chi}=\frac{6}{C^2}-1\,, \quad w_{\rm eff}=\frac{1}{3}C^2-1\,.
	\label{eqn:FPFfixed}
\end{equation}
The repulsive nature of fixed point ${\bf F}$ is illustrated in the left panel of Fig.~(\ref{fig:phasePortraitsQ2}). The only fixed point which admits a non--vanishing background fluid density with a sizeable contribution from both axion and modulus is ${\bf G}$. Here, the modulus energy density is dominated by its kinetic contribution, whist the axion remains frozen, its motion suppressed by Hubble friction. Both the axion and modulus track the evolution of the dominant background fluid
\begin{eqnarray}
	\rho_\phi&=&\frac{9H_i^2}{2C^2}\gb(2-\gb)\left(\frac{a}{a_i}\right)^{-3\gb}\,, \nonumber \\
	\rho_\chi&=&\frac{9H_i^2}{2C^2}\gb^2\left(\frac{a}{a_i}\right)^{-3\gb}\,,
	\label{eqn:FPGdensities}
\end{eqnarray}
whilst giving a background density $\Omega_{\rm b}=1-3\gamma_b/C^2$. This dynamical attractor is precisely the axion and modulus tracking behaviour that was described in~\cite{marsh2011}. It is in the combined equation of state 
\begin{equation}
	w_{\rm eff}=\frac{3\gb(\gb-1)}{C^2}\,,
	\label{eqn:FPGfixed}
\end{equation}
(rather than the individual equations of state) that we see tracking as $w_{\rm eff}$ tries to follow the equation of state of the dominant component. Tracking is finally destroyed as axion oscillations begin, which is the cosmic trigger event that restabilises the modulus. We show evolution into this fixed point in the right panel of Fig.~(\ref{fig:phasePortraitsQ2}).

%%%%%%%%%%%%%
%%%%%%%%%%%%%

\subsubsection*{Additional Comments on Fixed Point ${\bf G}$}

%%%%%%%%%%%%%
\begin{figure}
\centering
\includegraphics[scale=0.95]{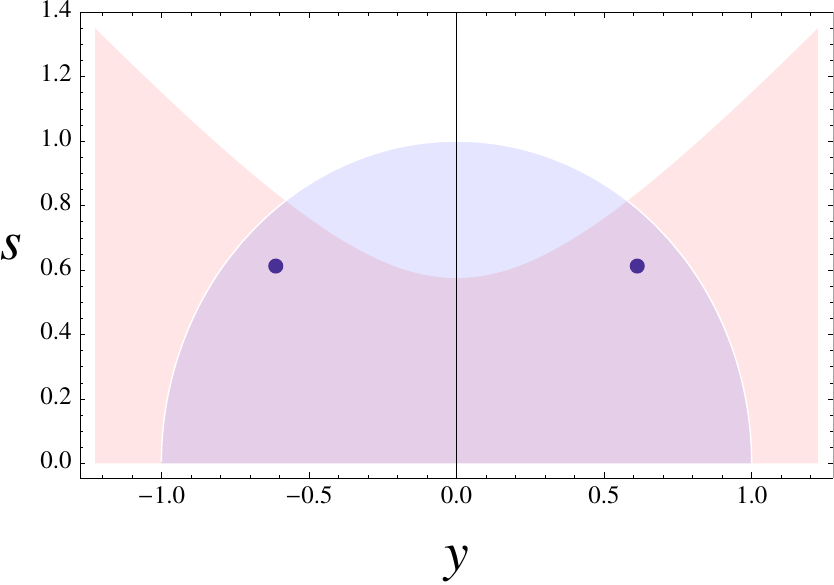}
\caption{Constraints relevant to trajectories approaching the quasi-stable fixed point ${\bf G}$ with a background matter fluid, $\gamma_b=1$. The blue shaded region under the semi-circle is the region allowed by the Friedmann constraint. The hyperbola bounds accelerated expansion, with the red shaded region having $w>-1/3$. The two points represent fixed point ${\bf G}$ for $s>0$. For $C>\sqrt{3}$ these lie inside the allowed region. We conclude that it is possible for trajectories approaching ${\bf G}$ to cross the $w=-1/3$ divide if they spiral as they do so, possibly leading to multiple epochs of accelerated expansion.}
\label{fig:g_bounds}
\end{figure}
%%%%%%%%%%%%%

As a particularly interesting fixed point, we choose to discuss some additional phenomenology relating to fixed point ${\bf G}$. Firstly we discuss accessibility of the fixed point. Even in the unbounded co-ordinates of an AdS minimum, approximate trapping in ${\bf G}$ requires variables other than $\{y,s\}$ to be  approximately zero, and so flatness bounds us with $y^2+s^2<1$, which defines a circle. This in turn imposes a constraint on $C$ as a function of $\gamma_b$ for ${\bf G}$ to be within this region:
\begin{equation}
C>\sqrt{3 \gamma_b}; \quad \text{physically accessible ${\bf G}$.}
\label{eqn:access_c}
\end{equation} 

An interesting phenomenon when entering fixed points in a multi-field model is the possibility of multiple periods of accelerated expansion \cite{catena2007}. When projected down to the $\{y,s \}$ subspace the condition for $w<-1/3$ in an expanding universe becomes $s>\sqrt{(\frac{2-\gamma_b}{\gamma_b})(y^2+\frac{\gamma_b-2/3}{2-\gamma_b})}$. 

A temporary trapping in point ${\bf B}$, where there is a larger value of cosmological constant than in the true vacuum, could lead to a single period of accelerated expansion during an otherwise radiation dominated era. In such a case, the axion and modulus fields would pick up large additional isocurvature fluctuations from this brief period of inflation. This period would end when the fields move towards their vacuum, and as such the global minimum would have to be dS.

The situation for ${\bf G}$ in this regard is more interesting. In a matter background, $\gamma_b=1$, this is pictured in Fig.~\ref{fig:g_bounds}. Here we see that it is possible to have both a flat universe and $w<-1/3$ in a $\{y,s\}$ dominated phase. Trajectories in the $\{y,s\}$ subspace will spiral into ${\bf G}$ if the eigenvalues of the stability matrix  ${\bf W}$ (that point in the $\{y,s\}$ directions) have an imaginary part. The stability analysis (see Appendix~(\ref{appdx:stability}), Eq.~(\ref{EigenG})) reveals that this is the case if:
\be
\label{eqn:multiple_acc_c}
C^3-8C^2+24>0\,.
\ee
For the parameter space of interest, $C>0$, this bound is satisfied for $C<2$ and $C>3+\sqrt{21}\approx7.6$. For any $C$ between these two values, the eigenvalues are real and trajectories will not spiral into ${\bf G}$ but move in straight lines, and so cannot cross $w<-1/3$. When Eq.~(\ref{eqn:multiple_acc_c}) is satisfied however, the trajectories in $\{y,s\}$ can spiral toward ${\bf G}$, having the possibility of crossing the $w<-1/3$ bound, perhaps multiple times. So, \emph{trajectories approaching this fixed point can lead to multiple periods of accelerated expansion during a matter dominated epoch}. This phenomenon is extremely tightly constrained: such an epoch of acceleration must be less than $0.05$ $e$--folds long \cite{linder2010}.

% -------------------------------------------------------------------------
\subsection{Scanning Parameter Space}\label{simulations}

Since the system of Eqs.~(\ref{eqn:autonomousSystem}) are first order and autonomous, they are very quick to integrate numerically. We exploit this nice property by performing a `scan' of the model parameter space around regions of interest, selecting particular scenarios to investigate more systematically. We use our scans to further our qualitative understanding of the phenomenology of the model and to locate and single out specific novel features.

The autonomous system has eight different parameters which determine the subsequent motion of any given trajectory in phase space: six initial conditions, $\{ x_{i},y_{i},z_{i},r_i,s_{i},t_{i}\}$ and two parameters, $\{C,\beta\}$. To ensure that this rather large parameter space is sampled in a uniform and efficient way, we use the method outlined in Appendix~(\ref{appdx:ICM}).

We briefly describe this process for initial conditions chosen to be close to fixed point ${\bf A}$, i.e. beginning in the fluid dominated phase with a non--vanishing background fluid density, $\Omega_{\rm b}(\text{initial})$. It is trivially generalised to the case of any other fixed point. For ${\bf A}$, with only some loss of generality, we make the simplifying assumption that the axion and modulus fields begin frozen, $x_i=y_i=0$. Then, using the Friedmann constraint, Eq.~(\ref{eqn:OmegaB}), and the vacuum constraint, Eq.~(\ref{eqn:defZeta}), we have
\be
\label{eq:manifold}
s_i^2=p-\zeta\frac{r_i^4}{z_i^2}+r_i^2-z_i^2\,,
\ee
initially. Here, $p=1-\Omega_{\rm b}(\text{initial})$. The initial conditions are constrained to lie on this three--dimensional manifold, which we will call $\mathcal{M}$. Scanning the initial conditions of the system then reduces to varying two initial conditions evenly over $\mathcal{M}$ with the third constrained by the equation for $\mathcal{M}$. We choose to vary $z_i$ and $r_i$, whilst still being free to independently vary $\{C,\beta\}$. Since $\mathcal{M}$ has non--constant curvature, it is not trivial to sample it in a uniform way and so we use a statistical sampling method which is presented in Appendix~(\ref{appdx:ICM}). Choosing a value of $p\approx0.01$ ($\Omega_{\rm b}(\text{initial})\approx 0.99$) is our definition of ``near'' to fixed point {\bf A}. 

Before we present the results of our numerical scans, it will further add to our intuition to briefly discuss the change in topology of the initial condition manifold either side of $\zeta=1/4$. If the vacuum is dS, $\zeta>\frac14$, the surface area of the manifold above some value $s=s_{\rm min}$ is finite. For $\zeta\le\frac14$, (AdS and Minkowski vacua) the manifold is not bounded above $s_{\rm min}$ and its surface area is infinite. Suppressing the subscript $i$ for brevity, this change in topology is best illustrated by solving Eq.~(\ref{eq:manifold}) at $s=s_{\rm min}$ for $z$. This generates two physically relevant roots as a function of $r$ which describe the curves where $\mathcal{M}$ intersects the $s=s_{\rm min}$ plane. These two curves meet at a point, $r_{\rm max}$,
\be
\label{eqn:r_max}
r_{\rm max}=\sqrt{\frac{(p-s^2_{\rm min})(1+2\sqrt{\zeta})} {4\zeta-1} }\,.
\ee
One can also obtain equations for the two $\mathcal{M}$-$s_{\rm min}$ intersection curves as a function of $z$. These two curves meet at 
\be
\label{eqn:z_max}
z_{\rm max}=\sqrt{\frac{4\zeta(p-s^2_{\rm min})}{4\zeta-1}}\,.
\ee
Eqs.~(\ref{eqn:r_max}) and~(\ref{eqn:z_max}) illustrate the change in topology of $\mathcal{M}$: for $\zeta=1/4$, $r_{\rm max}\,,z_{\rm max}\rightarrow\infty$, whilst for $\zeta<1/4$, $r_{\rm max}\,,z_{\rm max}\in\mathbb{C}$. In both cases, the manifold never intersects the $s=s_{\rm min}$ plane. Only for $\zeta>1/4$ is the surface area of the manifold bounded above $s_{\rm min}$. This is another clear example of the genuine change in the geometry of phase space when the vacuum is chosen to be either dS or AdS. \\

We now begin to discuss the results and findings of our numerical analysis. Our ability to perform scans of this kind has many possible applications for investigating the cosmological phenomenology of our model. Here we choose to simply show some examples that illustrate the capabilities of our technique. All models we present have a dust background fluid, $\gamma_b=1$ and a dS vacuum. We run two large simulations: \texttt{FP-A} and \texttt{FP-G}. For simulation \texttt{FP-A}, we evolve $562500$ models, each chosen to begin in fixed point ${\bf A}$ with $\Omega_{\rm b}(\text{initial})=0.99$. We scan $\{ C, \beta \}$ evenly in logarithmic space on a $25\times 25$ grid, and at each point we use our initial condition algorithm to evenly sample the space of $\{z_i,r_i,s_i \}$ over $\mathcal{M}$. For \texttt{FP-G}, we evolve $506100$ models, each chosen to begin near to fixed point ${\bf G}$. Unlike simulation \texttt{FP-A}, we are not free to independently vary the initial conditions and $C$ since in fixed point ${\bf G}$, $\Omega_{\rm b}=1-3/C^2$. Furthermore, $y_{\rm c}=\sqrt{\frac{3}{2}\frac1C}$ and so $y_i\neq0$. Hence, every time $C$ and $y_i$ are changed, the shape of the initial condition manifold also changes. Therefore we absorb $y_i$ and $\Omega_{\rm b}$ into the parameter $p$ of Eq.~(\ref{eq:manifold}): $p=\frac{3}{C^2}-y_i^2$. We then vary $C$ logarithmically and $y_i$ linearly across a $10\times 15$ grid and use our initial condition algorithm to evenly sample the space of $\{z_i,r_i,s_i \}$ over $\mathcal{M}$ for each point, $\{ C, y_i \}$, where $\mathcal{M}$ has a different shape. $\beta$ is varied $15$ times on a logarithmic scale.

In both simulations, individual models are terminated under two conditions: either they have settled into fixed point ${\bf M}$ for more than 5 $e$--folds, or, they have ran for a total of more than 500 $e$--folds. The results are presented so that at each point in $\{ C, \beta \}$ space, the average over all trajectories on $\mathcal{M}$ is taken, or alternatively for each point in $\{ z_i,r_i \}$ space we could average over parameters $\{C, \beta \}$, i.e. repeated points in any plane have their contour value averaged.

\subsubsection{The End of Fluid Domination}

\begin{figure}
\centering
\includegraphics[scale=0.37]{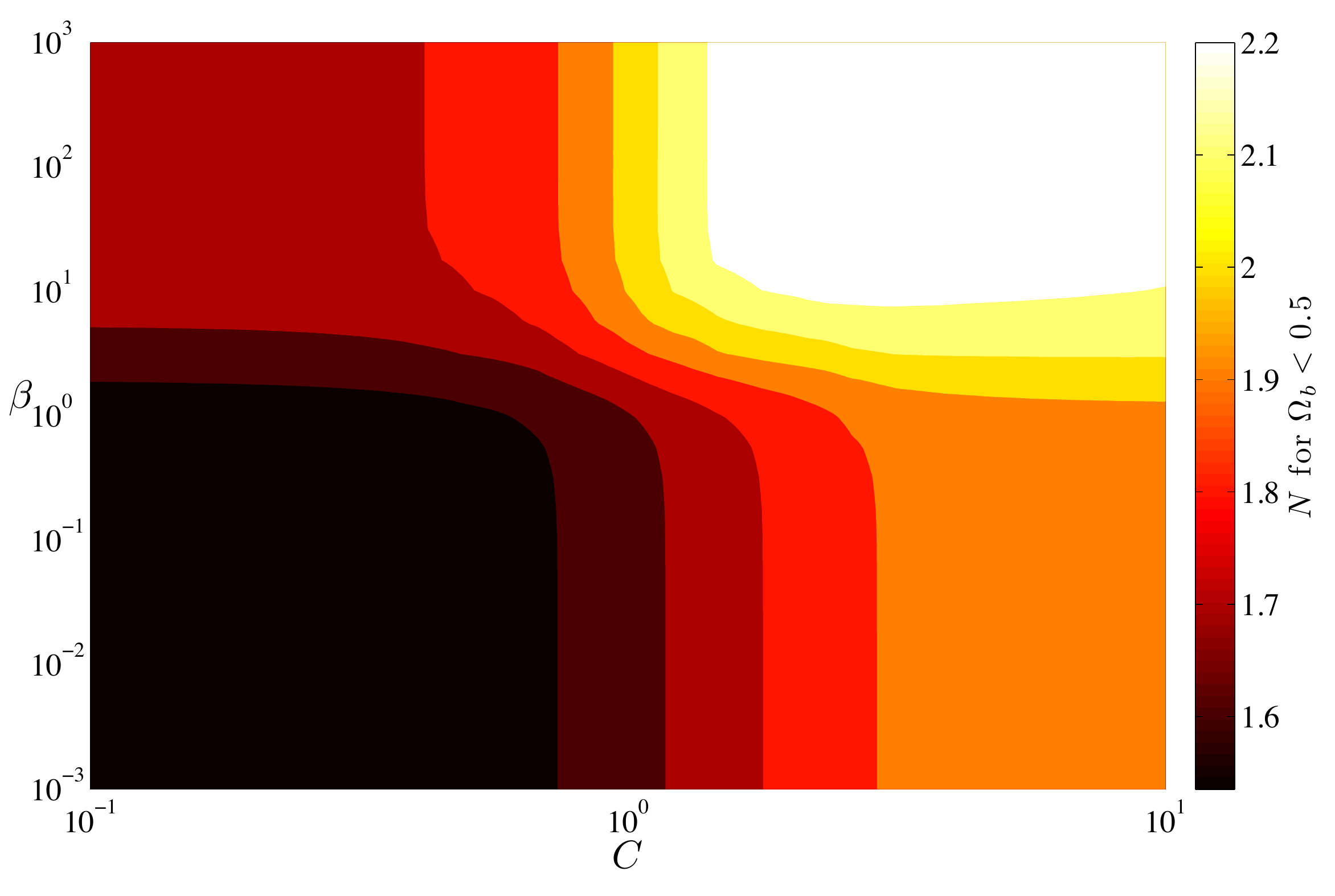}
\caption{Contour plot of the number of $e$--folds, $N_{\rm efd}$, when $\Omega_b<0.5$ for a scan over parameter space of all models beginning near fixed point ${\bf A}$ with $\Omega_b=0.99$ (the \texttt{FP-A} simulation). Each point in $\{ C,\beta \}$ represents an average over the initial condition manifold $\mathcal{M}$.}
\label{fig:c_beta_nFLUIDendA}
\end{figure}

In Fig.~(\ref{fig:c_beta_nFLUIDendA}) we plot in the $\{C,\beta \}$ plane, for models from the \texttt{FP-A} simulation, the number of $e$--folds, $N_{\rm efd}$, from the beginning of the evolution until the end of fluid domination when $\Omega_b<0.5$.

%%%
%%%

Before analysing the figure in more detail, it is first worth thinking about what we should expect from such a number. The maximum across all models occurred for $N_{\rm efd}\approx 5.3$ while the minimum occurred for $N_{\rm efd}\approx 1.5$. It is simple to show from the Friedmann equation that a $\Lambda$CDM cosmology beginning with $\Omega_{\rm m}(\text{initial})=0.9$ will reach $\Omega_m=\Omega_\Lambda=0.5$ after $N_{\rm efd}\approx 1.5$ $e$--folds. This is because, in the absence of energy input, a cosmological constant maximally decreases $(1-\Omega_m)$: this should be the limiting case of our model when the fields are frozen, which indeed it is. On the other hand, a model beginning at matter-radiation equality at $a_{eq}\approx 10^{-3}$ has $(1-\Omega_{\rm m}(\text{initial}))\sim \mathcal{O}(10^{-9})$, and depending on $\Omega_{m,0}$ has $6\lesssim N_{\rm efd}\lesssim 7$. 

In our model $N_{\rm efd}$ can be increased and approach this limiting case in three ways. The fields can oscillate before they overtake the fluid density, they will then scale like matter, always remaining sub-dominant and the end of fluid domination will be caused by the cosmological constant. Secondly, they could enter a scaling solution, where they also remain a fixed sub-dominant fraction of the energy density. Thirdly, they could roll to the minimum of the potential, reducing the vev due to the negative energy term in the modulus only part of the potential. We see that our maximum of $N_{\rm efd}$ approaches the limiting case, being slightly below it as some time is taken for these dynamics to occur.

In Fig.~(\ref{fig:c_beta_nFLUIDendA}) there is a clear correlation of $N_{\rm efd}$ with the parameters. Smaller average $N_{\rm efd}$ occurs for low $\beta$, where the axion mass is small preventing oscillations, and low $C$ where the scalar field energy density in scaling solutions is large. Larger average $N_{\rm efd}$ occurs for large $\beta$ and $C$ where oscillations can occur earlier and the energy density in scaling solutions is smaller.

When considering models from \texttt{FP-G} we imposed a cut for all $C<\sqrt{3}$, where ${\bf G}$ is unphysical (the initial conditions correspond to negative $\Omega_b$). The first difference observed from fixed point ${\bf A}$ was vertical cut giving very low $N_{\rm efd}$ at small $C$. These models had initially very small $\Omega_b$: the minimum of $N_{\rm efd}$ output from the code is $N_{\rm efd}=0.005$ which is our numerical step size in $N$, i.e. the models began out of fluid domination. Since $C$ sets the initial $\Omega_b$ for models beginning in ${\bf G}$, the general trend of increasing $N_{\rm efd}$ with $C$ continued and was dominant, until at large $C$ and $\beta$ it gave way to the effects described above in the case of ${\bf A}$. Axion oscillations lead to a decrease in $\phi$ to $\phi<\tilde{\phi}$ and therefore spoiled ${\bf G}$ after some short time, decreasing the overall scalar field density. In the case of an AdS negative potential minimum, these regions where ${\bf G}$ is spoiled would be those that eventually collapse.

\subsubsection{Multiple Periods of Accelerated Expansion}

Motivated by the fact that spiralling trajectories in phase space may generate multiple periods of accelerated expansion, we scan the model parameter space for this feature. We compute the number of periods of accelerated expansion, $\mathcal{N}_{\rm ae}$, by counting the number of times $w_{\rm total}=\sum_i w_i\Omega_i<-\frac13$ along the model trajectory. Here, $i$ labels the axion and modulus fields and the dust fluid. If one, or both scalar fields are oscillating about their minima, the \textit{averaged} equations of state, $\bar{w}_\phi$ and $\bar{w}_\chi$ are used in the calculation of $w_{\rm total}$. The average taken is a moving average and is re--calculated every $0.005$ $e$--folds as the trajectory advances in time. We define the onset of coherent oscillations as the time when the field velocity ($\dot{\phi}$ or $\dot{\chi}$) changes sign for the \textit{third} time. This ensures that we do not average any heavily or critically damped oscillations. This definition is somewhat arbitrary and so we should expect that $\mathcal{N}_{\rm ae}$ may be sensitive to the definition of the averaging process. Furthermore, for regions of parameter space where the fields are highly oscillatory (large $C$ and $\beta$), sampling the trajectory every $0.005$ $e$--folds may not be frequent enough to accurately average a single oscillation. We also note that this sampling rate is one tenth of the length of a period of accelerated expansion allowed by observation. Hence, computing $\mathcal{N}_{\rm ae}$ by taking the moving average of $w_\phi$ and $w_\chi$ is not always the observationally relevant procedure. 

With these limitations acknowledged, we consistently apply our definition of $\mathcal{N}_{\rm ae}$ to every single model in our simulations. From the \texttt{FP-A} simulation  we found that of our $562500$ models, $17668$ had $\mathcal{N}_{\rm ae}>1$. Of these, $676$ models were terminated after 500 $e$--folds for not reaching fixed point ${\bf M}$, so that the multiple $\mathcal{N}_{\rm ae}$ can be said to have definitely occurred near to a fixed point or the local minimum.  We also found 54 models with the largest $\mathcal{N}_{\rm ae}=8$. Of the remaining models with $\mathcal{N}_{\rm ae}=1$, $176599$ were terminated for not reaching ${\bf M}$ after 500 $e$--folds, and were thus still on the potential plateau trapped in ${\bf B}$. $\mathcal{N}_{\rm ae}$ may increase in future for these models, but the time scale is immense: situating them today, 500 $e$--folds gives $\Delta t = \Delta N/H_0\sim 10^{12-13}$ years\footnote{The time scale for collapse out of ${\bf B}$ or ${\bf G}$ for similar trajectories with an AdS minimum would be similar.}. We stress that we are not proposing any measure or figure of merit for fine tuning in this model, and as such the specific number of models pertaining to each case does not have any (clear) meaning. 

\begin{figure}
\centering
\includegraphics[scale=0.45]{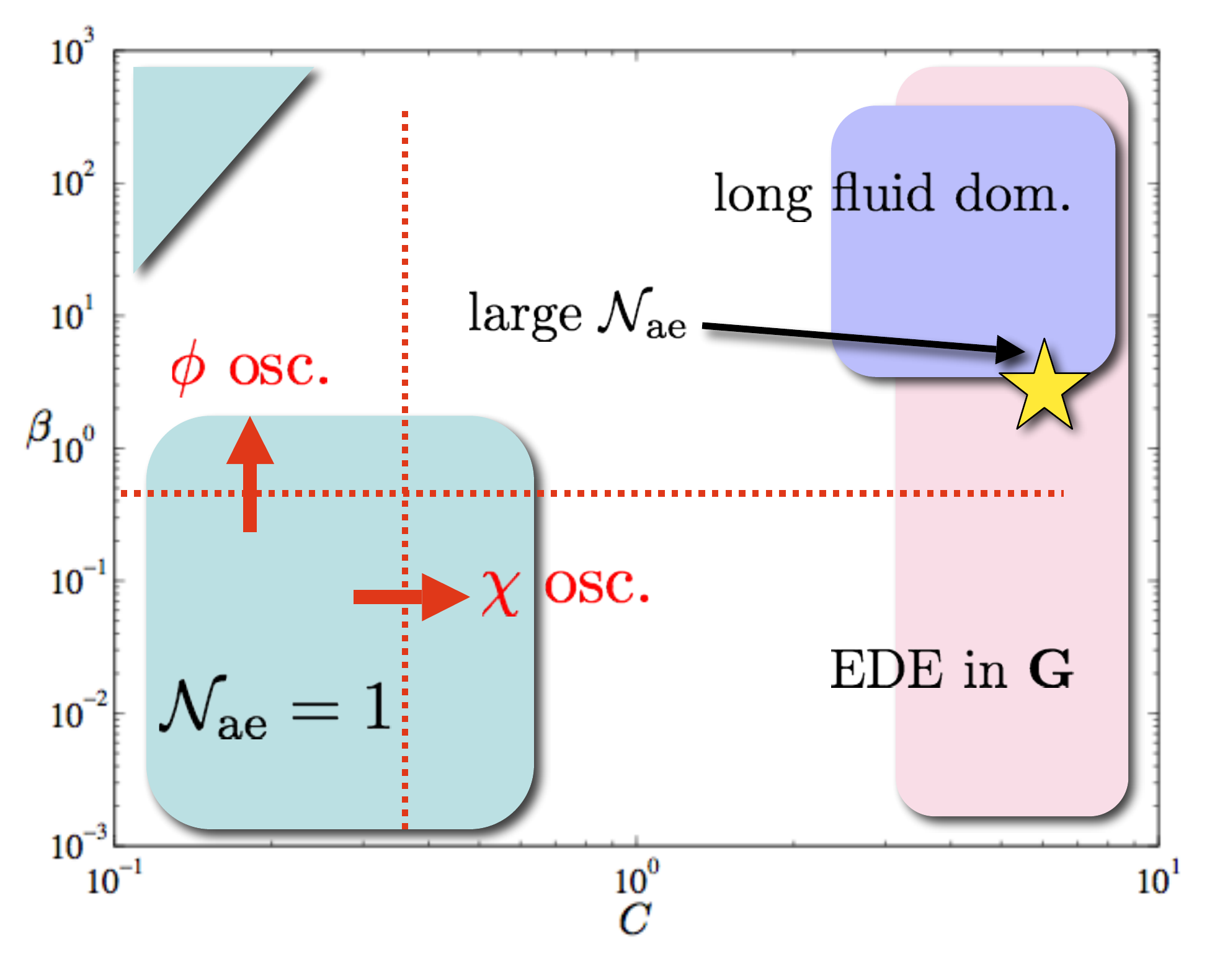}
\caption{Schematic summary of findings in $\{C,\beta\}$ space. $C$ is the exponent in the modulus potential, and gives the coupling between axion and modulus. $\beta$ is defined in Eq.~(\ref{eqn:beta}) and represents a ratio of scales between the axion and modulus terms in the potential. They are the only two parameters that appear in the dynamical system, Eqs.~(\ref{eqn:autonomousSystem}). The vertical and horizontal lines for field oscillations are the conditions of Eq.~(\ref{eqn:bifurcationMatCrossing}) for $\zeta\approx 0.3$ dS vacuum (see Eqs.~(\ref{eqn:defZeta}), (\ref{eqn:vacuum_cc})), and they move in the directions shown for increasing $\zeta$. The regions with $\mathcal{N}_{\rm ae}=1$ periods of accelerated expansion over all initial conditions only occurred for models from the \texttt{FP-A} simulation.}
\label{fig:c_beta_schematic}
\end{figure}

When considering the the distribution of $\mathcal{N}_{\rm ae}$ against $\{C,\beta\}$ we took all models with $\mathcal{N}_{\rm ae}>1$ and averaged over $\mathcal{M}$ as described above. We found some large regions of parameter space with $\mathcal{N}_{\rm ae}=1$ over all of $\mathcal{M}$. We also saw that there was a high density of large $\mathcal{N}_{\rm ae}$ at larger values of $C$ and intermediate values of $\beta$, with one clear peak. We show these locations schematically in Fig.~(\ref{fig:c_beta_schematic})

Our results also showed an interesting correlation between three dependent (output) variables where it was noticed that trajectories with large $\mathcal{N}_{\rm ae}$ occurred in those cosmologies that at the exit from fluid domination (entering the current epoch) contained only small values of $\Omega_\phi$ and $|\Omega_\chi|$ (it is consistent in this model to have $\Omega_\chi<0$ since it does not contain $\rho_\Lambda$: the total energy density remains always positive). This, combined with the larger values of $C$ in these regions, as we will discuss below, suggests that these models were likely in or near to ${\bf G}$ (or ${\bf B}$) at this time (again, see the schematic Fig.~(\ref{fig:c_beta_schematic})). Small values of $|\Omega_\chi|$ and $\Omega_\phi$ for light axions are those allowed by current data (we discuss some bounds in Section \label{discussion}), but is also potentially detectable with next generation experiments \cite{marsh2011b,calabrese2010}. Our scan suggests that such a cosmology could reasonably expect to undergo multiple periods of accelerated expansion in the future, and may have in its past. We re-state the bound from above: \cite{linder2010} showed that an intermediate epoch of accelerated expansion in the matter era must have lasted less than $0.05$ $e$--folds. 

Finally, our results showed that that almost all models with $\mathcal{N}_{\rm ae}>2$ had begun on trajectories with $\phi>\tilde{\phi}$, i.e. with a destabilised modulus, and thus access to ${\bf G}$ (see below). We reiterate that we have only analysed the dS case in this example: such allowable cosmologies may undergo a different cycle of $\mathcal{N}_{\rm ae}$ before collapse in the AdS case.

%%%
%%%
% -------------------------------------------------------------------------
\section{Discussion}\label{discussion}

\subsection{Phenomenology of Fixed Points}

The analysis in Section~(\ref{phasespace}) showed the existence of many fixed points, with various degrees of stability. Discussing the possible phenomenological implications of all of these would be a long and tedious process that we choose not to engage in. However, we will find it illuminating to discuss some properties of fixed point ${\bf G}$, both by way of example, and since we will find them to be particularly interesting.

The first thing to note about the fixed points is whether or not they occur in the bowl of the potential or on the plateau. We have that $s/r=\phi/\tilde{\phi}$, so that if $s/r>1$ the fixed point is on the plateau where the modulus is destabilised. For points ${\bf G}$ and ${\bf F}$ it is clear that they are on the plateau. Fixed point ${\bf B}$ has $s=r=0$, so the ratio is undefined, and is technically at infinite modulus value. Hence in \cite{marsh2011} temporary trapping was observed with large initial field values and a destabilised modulus, with exit from the fixed point occurring as the axion field value decayed. 

${\bf G}$ is the only fixed point that allows for scaling solutions where both axion and modulus track the dominant fluid component. This was the tracking behaviour observed in \cite{marsh2011}. This can be of particular use in \emph{alleviating fine tuning of axion initial misalignment angles} in the following way. Heavy axions require fine tuning of their initial misalignment angle if they are not to ``overclose'' the universe by causing matter-radiation equality to occur at too high a redshift: they are outside the ``anthropic window'' \cite{axiverse2009,marsh2010,mack2009a,mack2009b}. However, by allowing for tracking in the radiation era, the axion energy density will scale as $1/a^4$, instead of remaining a constant. Eventually oscillations will set in, since ${\bf G}$ is a saddle point, and the axion dark matter will scale as $1/a^3$, however this will begin from a lower energy density. The energy density is dumped into modulus kinetic energy of overdamped motion. This scaling will manifest as EDE, which we discuss in the next subsection. The difference to more standard tracking models is that the saddle point nature of ${\bf G}$ caused by the axion mass provides a natural mechanism for exit from tracking. Also in this model the tracking EDE field is not required to be the same as the field responsible for late time accelerated expansion, i.e. we have the additional $\rho_\Lambda$.

\subsection{Early Dark Energy}

Temporary trapping in ${\bf G}$ during the radiation era is phenomenologically attractive because, as also pointed out in \cite{marsh2011}, it has the possibility of leaving observable, and therefore constrainable, consequences as EDE. When is this situation possible? First, the modulus must be destabilised by axion initial misalignment, given by the bounds of Eqs.~(\ref{eqn:minbound}) and (\ref{eqn:shift_bound}). The modulus will roll out towards $\chi \rightarrow \infty$ until Hubble friction stops it (if in additon $C<\sqrt{6}$ then a temporary axion-modulus domination in ${\bf F}$ will occur). Later, if the bound of Eq.~(\ref{eqn:access_c}) is satisfied, tracking will begin. The effects of this will further bound $C$.

The axion-modulus EDE energy density contributes an amount $\Omega_e=\Omega_\phi+\Omega_\chi= 3\gamma_b/C^2$ during any period of tracking. EDE phenomenology places upper bounds on $\Omega_e$ that translate simply to bounds on $C$:
\begin{equation}
C>2 \Omega_e^{-1/2} \,.
\label{eqn:omegae}
\end{equation}

During the radiation era the scaling EDE will behave as an extra effective relativistic species, $\Delta N_{\rm eff}$, contributing to the background expansion. The density contribution can then be constrained by Big Bang Nucleosynthesis (BBN) and CMB bounds on $N_{\rm eff}$. For example, the BBN constraints of \cite{nollett2011} allow for $N_{\rm eff}=3.85\pm 0.26$, consistent with no change between BBN and the CMB. Taking the central value, parameterising the energy density as \cite{shvartsman1969,steigman1977}, and assuming all the additional energy density to be in the form of EDE allows for $\Omega_e \lesssim 0.1$:
\begin{equation}
C\gtrsim 6.2 \, \quad \text{BBN $\Delta N_{\rm eff}$ only, \cite{nollett2011}.}
\end{equation}

This large value of $\Omega_e$ would, however, be in conflict with the CMB (the agreement in \cite{nollett2011} was for $N_{\rm eff}$ only, and neutrinos behave differently in perturbations than EDE due to, for example, anisotropic stress). One of the main effects of the presence a sizeable $\Omega_e$ on the CMB is to change the location and amplitude of the acoustic peaks. The location of the first peak is related to the size of the sound horizon at decoupling which is given by
\be
r_s(a)=\int_{0}^{a} {\rm d}a\frac{{\rm d}\tau}{{\rm d}a}c_s\,.
\label{eqn:rs}
\ee
Here, $c_s^{-2}=3(1+R)$ is the sound speed of the photon--baryon fluid and $R(a)=\frac34\frac{\rho_b}{\rho_\gamma}$ is the photon to baryon ratio. Using the Friedmann Equation today (subscript $0$) and at an epoch during the radiation era when the universe has evolved to point ${\bf G}$ yields:
\be
\left(\frac{{\rm d}a}{{\rm d}\tau}\right)^2=H_0^2\left[\frac{\Omega_{m,0}a +\Omega_{\gamma,0}}{1-\Omega_e} \right]\,.
\ee
We will assume that the presence of \textit{two} background components, radiation, subscript $\gamma$, \textit{and} matter (dark and baryonic), subscript $m$, does not change the result $\Omega_e=\frac{4}{C^2}$ during the radiation dominated era.
Performing the integral in Eq.~(\ref{eqn:rs}) (similarly to~\cite{Hu1994Anisotropies}) from the last scattering surface (lss) to the epoch of matter radiation equality (eq)  gives:
\begin{align}
r_s=&\frac{4}{3H_0}\sqrt{1-\frac{4}{C^2}}\sqrt{\frac{\Omega_{\gamma,0}}{\Omega_{m,0}\Omega_{b,0}}}  \nonumber   \\ 
&\times {\rm ln}\,\left[\frac{ \sqrt{1+R_{\rm ls}} + \sqrt{R_{\rm ls} +R_{\rm eq}} }{1+\sqrt{R_{\rm eq}}} \right] = \sqrt{1-\frac{4}{C^2}}r_{s0}
\end{align}
where $r_{s0}$ is the standard sound horizon. The location off the first peak multipole is then:
\be
l_{\rm peak}\simeq\frac{2\pi}{r_sH_0}=\frac{C}{\sqrt{C^2-4}}l_0\,,
\ee 
where the standard peak multipole is:
\be
l_0=\frac{2\pi}{r_{s0}H_0}\simeq200\,.
\ee 
The qualitative behaviour is clear: for smaller $C$, i.e., for larger $\Omega_e$, the first peak occurs at a higher multipole. Ref. \cite{joudaki2012} performed simultaneous fits for $\Omega_e$ and neutrino species along with other extended cosmological parameter sets, and found maximum values for $\Omega_e$ at the 95\% confidence level of a few percent, with the absolute limit being dependent on priors about the DE ($w>-1$: no crossing of the phantom divide) or the neutrinos ($N_{\rm eff}>3$ from the standard model). The central value for $N_{\rm eff}$ in these fits was around $N_{\rm eff}\sim 3.6\pm 0.6$. Taking the most generous upper limit of $\Omega_e<0.042$ gives:
\begin{equation}
C\gtrsim 9.8 \, \quad \text{CMB and $\Delta N_{\rm eff}$, \cite{joudaki2012}.}
\end{equation}

Currently, there is no detection of $\Omega_e$, but it will be possible to detect with current and future CMB experiments of Planck and CMBPol \cite{calabrese2010}. Ref. \cite{calabrese2010} reports, for a fiducial Planck central value of $\Omega_e=0.03$ and marginalising over their other extended DE parameters, an error of $\sigma_{\Omega_e}=0.003$. A $3\sigma$ measurement of $\Omega_e$ translates to a bound:
\begin{equation}
10.1 \lesssim C\lesssim 13.8 \, \quad \text{Planck forecast, \cite{calabrese2010}.}
\end{equation}

It is worth noting finally that the CMB and BBN bounds need not both apply, since the fields do not have to have entered the scaling solution at any particular era, and can leave it. Of course there is also the caveat that these bounds only apply to the extent that motion in and near ${\bf G}$ is accurately described by the parameterisations used to derive them (\cite{joudaki2012} used a modified version of the parameterisation of \cite{doran2006}), and that approximately stable evolution in ${\bf G}$ can be maintained for long enough.

In Fig.~(\ref{fig:c_beta_schematic}) we show a schematic for the phenomenology in different regions of $\{C,\beta\}$ parameter space that the results of this discussion and Section~(\ref{simulations}) have led us to.

\subsection{The Assumption of Fixed $f_a$, and Uplifting the Potential}

Throughout this work, as we have mentioned, we have assumed that $f_a$ can be taken fixed and that the modulus only effects the axion through exponentially scaling the mass. This had the simplifying property of providing a trivial metric on field space, with no change to the canonical kinetic terms. We can look at the validity of this assumption by computing what the change in $f_a$ would be along any particular trajectory. The assumption will be approximately valid if:
\begin{equation}
|\Delta f_a |/ f_{a,0} \lesssim 1 \,,
\end{equation} 
where $f_{a,0}$ is the point on the trajectory deemed to be ``today'' and the difference is calculated from the last relevant epoch. In the axiverse the scale of $f_a$ is fixed around $10^{16}\unit{GeV}$ by fixing the product $S=C\chi \sim 200$. This does not appear in our dynamical systems analysis, since the scale of $\chi$ only comes in via $\omega$, which the system does not depend on.

As mentioned in Section \ref{potential} the change in $f_a$ will most likely be large for any trajectories that begin on the plateau of the potential and end in the bowl. This would require us to compute corrections in moving, for example, \emph{between} fixed points ${\bf B}$, ${\bf G}$ and the global minimum ${\bf M}$. In the vicinity of the fixed points, checking that $f_a$ remains roughly fixed would require specifying $\omega$ and checking on a case-by-case basis.

It is possible that trapping in ${\bf G}$, or any fixed point with non-zero $y$, for an extended period of time could lead to large $\Delta f_a / f_a$. We can estimate this effect as follows. For a trapping of $\Delta N$ $e$--folds in ${\bf G}$ and setting $C\chi_i = 200$ to get the correct $f_a$ for the axiverse in the early universe, factors of $C$ cancel and we have:
\begin{equation}
\frac{\Delta \chi}{\chi_i}=-\frac{\Delta f_a}{f_{a,0}} = \frac{3 \gamma_b}{200}\Delta N = \mathcal{O}(10^{-2}) \Delta N \, .
\end{equation}
This will always be small for any scenarios of interest, since $\Delta N$ could only be large if ${\bf G}$ were driving inflation but we have seen that ${\bf G}$ itself cannot be accelerated and hence this is impossible. $f_a$ today will be only $\Delta N$\% away from its initial value for small $\Delta N$, and therefore predictions based on trapping in ${\bf G}$ in any particular epoch should be unaffected by our assumption of fixed $f_a$. However, in predicting the fate of the universe, we emphasise again that ${\bf G}$ is unstable and moving into ${\bf M}$ in the future (or in the current epoch, as in \cite{amin2011}) may entail large changes in $f_a$.

While the dynamical effect of changing the kinetic terms is hard to predict, it is simple to compute the change in the potential caused by identifying $f_a = \frac{1}{C \chi}$. The coupling term in the potential becomes:
\begin{equation}
U(\phi,\chi) = \frac{\mu^4}{2} C^2 e^{-C \chi}\chi^2 \phi^2\, . 
\end{equation}
This has one very interesting property: the emergence of a new, meta-stable (in the sense that it has a small barrier that can be tunnelled through, like the potentials of \cite{kachru2003}) modulus minimum in the region of large $\phi$. This meta-stable minimum can have \emph{positive cosmological constant, with no need for additional uplifting}, i.e. with $\rho_\Lambda=0$. However, it is unstable in the axion direction, and could only usefully drive current accelerated expansion with an ultra-light axion of mass $m_a \lesssim 10^{-33}\unit{eV}$. The emergence of the new minimum at large $\phi$ can be traced to the extra term in $\partial_\chi U$ with opposite sign. 

For the new minimum to emerge one requires $\chi < 2/C$ which makes $S\sim \mathcal{O}(1)$ and pushes $f_a \rightarrow M_{pl}$. This leads to more fine tuning if this minimum is to provide late time acceleration since the small axion mass necessary for stability  then needs to be put in by hand from the non-perturbative side, ruining the naturalness of the axiverse scenario for light axions. We leave further study of the properties of this $f_a (\chi)$ scenario, particularly its possibility of giving an alternative axion inflationary model, to a future work.

% -------------------------------------------------------------------------
\section{Conclusions}\label{conclusions}

In this paper we have studied a rich model of the dark sector, with many possible observational signatures as Dark Matter and Dark Energy, that extends and builds on well known work and tries to bring it into a broader theoretical context. Some of our findings are summarised in the schematic of Fig.~(\ref{fig:c_beta_schematic}).

Axions and moduli are intimately linked to the problem of the cosmological constant. Polchinski argued some time ago \cite{polchinski2006}, and indeed it has been known since the earliest days of string theory \cite{witten1984,schwarz1992} that the lightness and profusion of axions is a natural consequence of the theory, and is related to the anthropic demand for a small cosmological constant. Ultra-light axion fields with the hierarchy of masses generated by exponential dependence on the internal geometry of the compact space are observationally relevant as a distinct form of dark matter. We have studied the cosmological evolution of axions when the energy scale of the potential is allowed to be dynamically controlled by a modulus of this geometry, instead of remaining fixed. If both the potential of the axion and the modulus arise from non-pertrubative physics at similar energy scales, then we have shown that axion initial misalignment can leave the modulus destabilised in the early universe and when the axion is allowed to be cosmologically light this can lead to significant evolution of the modulus throughout cosmic history. This evolution allows for the possibility that the vacuum energy can change sign through the course of cosmic evolution, with today's quasi de-Sitter expansion being only temporary and the future evolution of the universe is a Big Crunch cosmic doomsday. We have demonstrated the topological change in the allowed phase space that such a possibility creates in two separate coordinate systems.

By looking for fixed points of the dynamical system we have shown that a modulus destabilised by a large axion initial misalignment gives rise to a scaling solution where excess energy density in the axion field is redshifted away during the radiation era. One can always argue that a valid model must cut off the fine tuning on axion fields to values where there is a stable modulus, and we have shown the ratios of scales necessary for this. Alternatively, if destabilisation does occur and such a tracking behaviour ensues during cosmic evolution, then fine tuning on the axion dark matter is alleviated and in addition observational limits on Early Dark Energy place constraints on the couplings of the model.

We have not discussed the possibility of fitting this model to be cosmologically viable, which would require fixing $H_0$ and $w(z)$, among other things. Fits of this kind would allow comparison to current and projected constraints on $w(z)$ and distance measurements (see, e.g. \cite{copeland2006,euclidRB,euclidWEB}), and in the context of this model trapped in ${\bf B}$ or ${\bf G}$ would allow predictions for future vacuum decay to dS in ${\bf M}$ \cite{amin2011} or collapse \cite{kallosh2003,wang2004}. It is worth noting, however, that even small uncertainties in the curvature, $\Omega_k$, can produce significant degeneracies and misestimations of $w(z)$ from distance measurements alone \cite{clarkson2007,hlozek2008}, which highlights the need for more complete models, and use of more experimental estimators, when discussing non-standard models of Dark Energy. We have also not discussed perturbations, which would be necessary to compare this model properly to large scale structure or CMB measurements.

If the modulus in this model controlled a coupling of the standard model, then predictivity of any model building will demand for it to be stabilised, and observational constraints will demand variations caused by axion evolution to be small, although potentially observable (see, e.g., \cite{lee2004,calabrese2011})\footnote{Many analyses of this kind, however, fail to account for the huge effect that variation in $\alpha$ would have on the standard model contribution to the vacuum energy through vacuum bubbles, at best greatly worsening fine tuning, and at worst ruining most anthropic explanations for the smallness of $\Lambda$.}. Our analysis showed that this would lead to a tuning on axion initial misalignment, in addition to any related to dark matter density, if $\phi_i < \tilde{\phi}$. If this bound is violated and the modulus is destabilised, any low energy constants that depend on it will be stabilised by Hubble friction and  eventually scale according to the dynamics of a fixed point. We have shown that trapping in such a fixed point can reasonably maintain the axion decay constant, $f_a$, and so may also be expected to naturally maintain any other constant with similar modular dependence. Anthropically, the meta-stability of this state of affairs is only as unnatural as a generic model allowing for a future Big Crunch.

The future singularity allowed in the parameter space of this model changes the asymptotic structure of spacetime and may be relevant to holographic models, or ``Cosmology/CFT'' \cite{strominger2001,maldacena2010,harlow2010,kanno2011}, although a rolling rather than tunnelling to an AdS state in our model may trivialise any specific holographic mapping. Allowing for long lived unstable scalar potentials muddies the waters somewhat in the question of fine tuning in the landscape. The axiverse and supergravity \cite{kallosh2002,kallosh2002b,kallosh2003,kallosh2003b} naturally allow for scalar masses around $H_0$, but string quintessence models run up against many problems \cite{kaloper2009}, although for axions successful models do exist \cite{panda2010,kim2003,kim2009}\footnote{During the final stages of preparation of this manuscript a very interesting model for natural, and indeed coupled, quintessence in string theory was proposed in \cite{cicoli2012b}. In particular, this involved a modulus controlling the size of a four-cycle, which in Type IIB theory can have a $C_4$ axion associated to it. The mass of this axion will depend on the quintessence field, realising our model. In that work, the important constraints of fifth-force experiments and SUSY breaking are also addressed.}. However, if the landscape favours instabilities \cite{marsh2011c} and as we have said they appear to be necessary feature in eternal inflation, it certainly seems pertinent to study their cosmology. Could it be that the seemingly unlikely situation of many light axions pulling the moduli hither and thither in ultimately collapsing universes in fact opens up a whole new part of the ``wasteland'', or that axion friction can favour a large number of destabilised moduli and a natural route to non-trivial quintessence? What types of universe dominate the (admittedly controversially-defined) landscape volume: unstable, cosmological constant, cyclic or quintessence? We have also seen that a coupling of axions and moduli can allow for large variations of the cosmological constant in the future, making multiple epochs of accelerated expansion possible during the matter dominated epoch in our past (observationally tightly constrained by \cite{linder2010}), or in the future evolution of the universe. Does this too effect our perception of fine tuning in relation to Dark Energy?

In closing, we like to hope that the study of this model will motivate string theorists to further consider late time effects that the existence of ultra-light axions can have on diverse aspects of string cosmology, and demonstrate a new and rich model in the dark sector to cosmologists.

%\vspace{-10pt}

% ------------------------ ACKNOWLEDGEMENTS ----------------------------------

\section*{Acknowledgements}

ERMT is supported by the University of Nottingham, and would like to thank Shuang-Yong Zhou, Francisco G. Pedro and Paul Saffin for useful discussions. PGF acknowledges the support of the Oxford Martin School and the Beecroft Institute for Particle Astrophysics and Cosmology. EJC would like to thank the STFC, the Leverhulme Trust and the Royal Society for financial support. DJEM acknowledges the support of an STFC studentship and would like to thank: Celia Escamilla-Rivera for useful discussions about dynamical systems in cosmology; Mustafa Amin and Subodh Patil, for insights about scalar fields and effective field theory; Sergei Dubovsky, Francisco G. Pedro, John March-Russell, Nemanja Kaloper, Paul Steinhardt and Juan Maldacena for useful discussions about cosmology and the landscape; and finally David Spergel and Princeton University Astrophysics for hospitality while part of this work was completed.
\vspace{-10pt}

\appendix

\section{Stability Analysis}\label{appdx:stability}

In order to study the stability of the fixed points we expand about these points, setting ${\bf X=X_c+\delta X}$, with ${\bf \delta X}$ the perturbations of the compact variables defined by Eqs.~(\ref{eqn:variables}) considered as a column vector. To first order, the perturbations satisfy ${\bf \delta X'=W\cdot \delta X}$, where the matrix ${\bf W}$ contains the coefficients of the perturbation equations. The general solution for the evolution of these linear perturbations can be written as 
\begin{equation}
	 {\bf X}=a_1{\bf q}_1e^{\lambda_1N}+\cdots+a_6{\bf q}_6e^{\lambda_6N}  \,,
	\label{generalSol}	
\end{equation}
where the ${\bf q}_n$ are the eigenvectors associated with the eigenvalues $\lambda_n$ of the matrix ${\bf W}$. Thus, the stability of the fixed points depends upon the nature of the eigenvalues. The eigenvector of the corresponding eigenvalue determines the directions in phase space with which the eigenvalue is associated. We use the following classification~\cite{Kibble2004Classical}
\begin{itemize}
	\item (i) Stable node: $\lambda_n<0$ for $n=1,...,6$
	\item (ii) Unstable node: $\lambda_n>0$ for $n=1,...,6$
	\item (iii) Saddle point: $\lambda_n<0$, $\lambda_m>0$, (or $\lambda_n>0$, $\lambda_m<0$) for $n,m=1,...,6$ with $n\neq m$
	\item (iv) Stable spiral: The determinant of the matrix ${\bf W}$ is negative and the real parts of the $\lambda_n$ are negative
\end{itemize}
For an expanding universe, a fixed point is an attractor (stable) in the cases (i) and (iv), but it is not so in the cases (ii) and (iii) (unstable). We use the notation $\lambda_n\{i,\cdots\}$ for an eigenvalue $\lambda$. The subscript $n$ labels the eigenvalue and the $\{i,\cdots\}$ denote the direction(s) in phase space with which this eigenvalue is associated, which may be determined by the $n^{\rm th}$ eigenvector.

The positive roots of the fixed points $z_c,r_c,s_c,t_c$ given in Table~(\ref{tab:fixedPoints}) lie in the expanding universe branch ($H^+$), for which $z,r,s,t>0$. The negative roots lie in the collapsing universe branch ($H^-$), where $z,r,s,t<0$.  For a collapsing universe, a fixed point is stable if the eigenvalues of ${\bf W}$ are \textit{positive}. This is because the `time' variable $N\equiv{\rm ln}\,(a)$ of the autonomous system~(\ref{eqn:autonomousSystem}) becomes a decreasing function of time.

The stability of the fixed points may be summarised as follows:

\begin{list}{\labelitemi}{\leftmargin=1em}
	\item {\bf Point A}
This is the trivial solution corresponding to fluid domination where the kinetic and potential components of the axion and modulus fields plus $\Lambda$ are negligible. It exists for all $C$, $\beta$ and $\gb$. In $H^+$, the eigenvalues are
\begin{align}
	\lambda_1\{x\}&=\lambda_2\{y\}        =-\frac{3}{2}(2-\gb)\,, \nonumber \\
         \lambda_3\{z\}&=\lambda_4\{r\}=\lambda_5\{s\}=\lambda_6\{t\} =\frac{3}{2}\,,
	\label{EigenA}	
\end{align}
so this is a saddle point in the full phase space. Point {\bf A} is unstable in the subspace of the $\{z,r,s,t\}$ directions and stable in the $\{x,y\}$ subspace. In $H^-$ this point remains unstable in the full phase space, since for realistic fluids $\gb<2$.
	\item {\bf Point B} 
The $\rho_\Lambda$ dominated asymptotic fixed point. In $H^+$, the eigenvalues are
\begin{align}
	\lambda_1\{x\}&=\lambda_2\{y\}         =-3\,, \nonumber \\
         \lambda_3\{t\}                  &=-3\gb\,, \nonumber \\
         \lambda_4\{z\}&=\lambda_5\{r\}=\lambda_6\{s\}=0  \,.
	\label{EigenB}
\end{align}
Notice this point has three \textit{zero} eigenvalues in the $\{z,r,s\}$ subspace: We say this is a \textit{marginally stable} solution in the sense that there is no instability growing exponentially, although it could be unstable to higher orders in the perturbation. To obtain the strict stability of this solution we would have to go beyond linear order, which we do not pursue as we know this point is ultimately unstable (see Section~(\ref{fixedpoints})). 
	\item {\bf Point C}
Corresponds to the axion and modulus kinetic dominated solution. For this point to exist, $-1\le y \le 1$. Rather than having an isolated fixed point, point ${\bf C}$ is formed of a continuous line of fixed points, known as an \textit{equilibrium manifold}, which we call a \textit{critical line}. This critical line is the unit circle $x_c^2+y_c^2=1$ and is a symmetry of the autonomous system with $z_{\rm c}=r_{\rm c}=s_{\rm c}=t_{\rm c}=0$. In both $H^+$ and $H^-$ the eigenvalues are
	\begin{align}
		\lambda_1\{z\}    &= 3\pm\sqrt{6}Cy\,, 		&\lambda_2\{x,y\}&=6-3\gb\,, \nonumber \\
		\lambda_3\{t\}    &=3\,,	&\lambda_4\{x,y\} &=0\,, \nonumber \\
		\lambda_{5}\{s\}&=\lambda_6\{0\} =3\pm\sqrt{\frac{3}{2}}Cy\,,
	\label{EigenC}
	\end{align}
where the $\pm$ in $\lambda_1$ and $\lambda_{5,6}$ corresponds to the $\pm$ of $y_{\rm c}=\pm y$. In General, if a nonlinear system has a critical line, the Jacobian matrix of the linearised system at a fixed point on the line has a zero eigenvalue with an associated eigenvector tangent to the critical line at the chosen point. Here, the zero eigenvalue $\lambda_4$ lies in  a direction tangent to the $x_c^2+y_c^2=1$ unit circle. This kind of non--linear system is a special subclass of the non--hyperbolic system (whose linearised system has one or more eigenvalues with zero real parts). 

The stability of a particular fixed point on the line can be determined by the non--zero eigenvalues, because near this fixed point there is essentially no dynamics along the critical line (i.e., along the direction of the eigenvector associated with the zero eigenvalue), so the dynamics near this fixed point may be viewed in a reduced phase space obtained by suppressing the zero eigenvalue direction. Then in $H^+$, this is point is unstable in the full phase space since $\lambda_3>0$.  

In $H^-$, where stability corresponds to positive eigenvalues, it might appear that stability is only guaranteed for certain values of $y$ along the fixed line. This is not so however, since both $y_{\rm c}=+y$ and $y_{\rm c}=-y$ (with $x_{\rm c}=\pm\sqrt{1-y^2}$) are fixed point solutions in a collapsing universe and correspond to the upper and lower halves of the unit circle respectively. This guarantees that every point along the critical line is stable. Hence, {\bf C} is the asymptotic future of any model with an AdS vacuum. The particular fixed point along {\bf C} that the system will finally evolve to will depend upon the initial conditions.
 
We also note that $\lambda_{5,6}$ (which has a multiplicity of $2$) is associated with only a single eigenvector, pointing in the $s$ direction: the matrix of linearised perturbation coefficients is \textit{defective}, i.e., it does not have a complete basis of eigenvectors, and is therefore not diagonalizable. As a result, the $r$ direction is not represented and we so we write $\lambda_6\{0\}$. This is almost certainly a consequence of the autonomous system having a higher dimensionality than required.
	\item {\bf Point D} 
Modulus dominated fixed point, where $V_{B}\gg V_{D}$. This points exists for $C\le\sqrt{\frac{3}{2}}$. In $H^+$ the eigenvalues are
	\begin{align}
		\lambda_1\{y,z\}    &= 4C^2-3\gb\,, 	& \lambda_2\{y,z\}    &=2C^2-3\,,  \,, \nonumber \\
		\lambda_3\{x\}    &=2C^2-3\,,	         &\lambda_4\{t\}         &=2C^2\,, \nonumber \\
	        \lambda_{5}\{r\}&=\lambda_5\{s\} =C^2\,.
	\label{EigenD}
	\end{align}
To be stable along the $\lambda_1$, $\lambda_{2,3}$ directions, which is the subspace $\{x,y,z\}$, requires $C<\sqrt{\frac{3}{4}\gb}$ and $C<\sqrt{\frac{3}{2}}$ respectively. The $\{r,s,t\}$ subspace is always unstable and hence point {\bf D} is unstable in the full phase space. Point {\bf D} remains unstable in the full phase space of $H^-$.
	\item {\bf Point E} 
A scaling solution where then axion energy density vanishes and the modulus energy density scales with the dominant background fluid with $V_{B}\gg V_{D}$. This point exists for $\gb\le2$. In $H^+$ the eigenvalues are:
	\begin{align*}
		\lambda_{1,2}\{y,z\}&= \frac{3}{4C}\bigg[ C(\gb-2)  \nonumber   \\ 
			&\pm\sqrt{(\gb-2)(9C^2\gb-6\gb^2-2C^2)} \bigg]\,,  \nonumber \\ 
	\end{align*}
	\begin{align}
		\lambda_3\{x\}&=-\frac{3}{2}(2-\gb)\,,  &\lambda_4\{t\}  &=\frac{3}{2}\gb\,, \nonumber \\
		\lambda_{5}\{r\}&=\lambda_{6}\{s\} =\frac{3}{4}\gb \,, 
	\label{EigenEplus}
	\end{align}
where the $+$root in $\lambda_{1,2} $ is for $\lambda_1$ and the $-$root is for $\lambda_2$. For $\lambda_{1,2}$ $\in\mathbb{R}$, $C\leq\sqrt{\frac{6\gb^3-12\gb^2}{9\gb^2-20\gb+4}}$. Furthermore, if this condition is satisfied, $\lambda_2$ is negative $\forall\, C,\gb$ and $\lambda_1$ is negative if $C>\sqrt{\frac{6\gb}{8}}$. When $\lambda_{1,2}$ has a imaginary part, the real part is always negative. The $\{r,s,t\}$ directions are always unstable and $\{x\}$ is stable for realistic $\gb$.  We see that point {\bf E} is unstable in the full phase space and remains unstable in the full phase space of $H^-$.
	\item {\bf Point F} 
Fixed point dominated by the modulus kinetic energy and potential energy of the axion--modulus coupling. This point exists for $C\le\sqrt{6}$. In $H^+$ the eigenvalues are 
	\begin{widetext}
	\begin{align*}
		\lambda_{1,2}\{y,s\}& =\frac{3}{2}\left[\frac{1}{2}C^2-\gb-1\right] \\
		&\pm\frac{1}{4}\sqrt{9C^4-36C^2+36C^2\gb+36 -72\gb+36\gb^2-8C^4\gb-8C^3+8C^3\gb+48C-48C\gb}  \,,
	\end{align*}
	\end{widetext}	
	\begin{align}
		\lambda_3\{x\}    &= \frac{1}{2}C^2-3\,, 		&\lambda_4\{t\} &=\frac{1}{2}C^2\,, \\ \nonumber 
		\lambda_5\{x,r\}  &=0\,,	                                    &\lambda_6\{z\} &=-\frac{1}{2}C^2 \,,
	\label{EigenF}
	\end{align}
where the $+$root in $\lambda_{1,2} $ is for $\lambda_1$ and the $-$root is for $\lambda_2$. We see immediately that point {\bf F} is unstable in the full phase space. In the presence of radiation, ($\gb=4/3$) there is a very finely tuned region of parameter space open to $C$ for which $\lambda_1$ and $\lambda_2$ may be real and negative. In the presence of dust, ($\gb=1$) the quantity under the square root simplifies to $C^4$ and the conditions for which $\lambda_1<0$ and $\lambda_2<0$ are $C<\sqrt{3}$ and $C<\sqrt{6}$ respectively. In $H^-$, point {\bf F} is always unstable in the full phase space.
	\item {\bf Point G} A scaling solution where the axion and modulus track the dominant background fluid. This point exists for $\gb\le2$. In $H^+$, the eigenvalues are:
	\begin{widetext}
	\begin{equation}
		\lambda_{1,2}\{y,s\} =\frac{3}{4C^2}\left[C^2(\gb-2)\pm\sqrt{C(\gb-2)(24\gb^3-24\gb^3C-48\gb^2+24C\gb^2+8C^2\gb+C^3\gb-2C^3)} \right] \,, \nonumber
	\end{equation}
	\end{widetext}	
	\begin{align}
		\lambda_3\{z\} &= -\frac{3}{2}\gb\,,  &\lambda_4(x)&=\frac{3}{2}\gb -3\,, \nonumber \\
		\lambda_5(t)  &=\frac{3}{2}\gb\,,  &\lambda_6(x,r) &=0 \,,
	\label{EigenG}
	\end{align}
where the $+$root in $\lambda_{1,2}$ is for $\lambda_1$ and the $-$root is for $\lambda_2$. For an expanding universe, there is a small region of $C$ parameter space for both dust and radiation where $\lambda_{1,2}$ are purely real and negative and a large region of $C$ where $\lambda_{1,2}$ has a complex part with the real part negative. In these regions, $\{y,s\}$ are stable directions. $\lambda_{5}$ is always positive however and so {\bf I} is unstable in the full phase space of $H^+$. In $H^-$, $\lambda_{3,4}$ are always negative and so {\bf I} is unstable in the full phase space of $H^-$.
	\item {\bf Point I} 
Modulus dominated fixed point, where $V_{D}\gg V_{B}$. Notice that this point mirrors point ${\bf F}$ and exists for $C\ge\sqrt{6}$. In $H^+$ the eigenvalues are:
	\begin{equation}
		\lambda_{3,4}\{x,s\} =\frac{1}{4}C^2-\frac{3}{2}\pm\frac{\sqrt{3}}{12}\sqrt{(C^2-6)(3C^2-18-8\beta^2)} \,,
	\label{EigenIplus}
	\end{equation}
	\begin{align}
		\lambda_1\{y,r\} &=C^2-3\gb \,, &\lambda_2\{y,r\}&=\frac{1}{2}C^2-3\,, \nonumber \\
		\lambda_5\{t\}&=\frac{1}{2}C^2\,, &   \lambda_6\{z\} &=-\frac{1}{2}C^2\,, 
	\end{align}
where the $+$root in $\lambda_{3,4} $ is for $\lambda_3$ and the $-$root is for $\lambda_4$. In $H^+$ regardless of the value of $\beta$, the real part of $\lambda_3$ can only be positive. These constrains are due to the existence condition $C\ge\sqrt{6}$. Hence, the $\{x,s\}$ subspace is always unstable. The conditions for stability for $\lambda_{1,2}$ in the expanding case are $C<\sqrt{3\gb}$ and $C<\sqrt{6}$, neither of which are ever satisfied if this point is to exist and so $\{y,r\}$ are unstable directions. We see that for both $H^+$ and $H^-$, point {\bf I} is unstable in the full phase space.
	\item {\bf Point M} 
The critical line in the $\{z,r,t\}$ subspace corresponding to the global axion--modulus minimum. The emergence of this critical line is due to the fact that the autonomous system is of one dimension too many. With the minimally required number of autonomous system variables, the line would degenerate to a unique point, given by Eq~(\ref{eqn:zetaCrossing}). {\bf M} exists for all $C\,,\gb\,,\beta$. It is important to point out that {\bf M} is only a fixed point in the presence of a dS vacuum ($\zeta>1/4$) and corresponds to the asymptotic future. If the vacuum is AdS ($\zeta<1/4$) the global minimum is not a fixed point, and the asymptotic future is cosmic doomsday in a Big Crunch. In $H^+$ the eigenvalues are: 
	\begin{align}
		\lambda_{1,2}\{y,z,r\} &= -\frac{3}{2}\pm\frac{1}{2}\sqrt{9-24C^2z^2}  \,, \nonumber \\
		\lambda_3\{z,r,t\} &=-3\gb\,, \nonumber \\
		\lambda_{4,5}\{x,s\}    &= -\frac{3}{2}\pm\frac{1}{2}\sqrt{9-8\beta^2z^2} \,, \nonumber \\
		\lambda_6\{z,r,t\} &=0\,.
	\label{EigenM}
	\end{align}
where the $+$root in $\lambda_{1,2} $ is for $\lambda_1$ and the $-$root is for $\lambda_2$. The $+$root in $\lambda_{4,5} $ is for $\lambda_4$ and the $-$root is for $\lambda_5$. Here, the eigenvector associated with $\lambda_6=0$ points in a direction tangent to the fixed line and so the stability of a particular fixed point on the line can be determined by the non–-zero eigenvalues, since near this fixed point there is essentially no dynamics along the fixed line (i.e., along the direction of the eigenvector associated with the zero eigenvalue) and the dynamics near this fixed point may be viewed in a reduced phase space obtained by suppressing the zero eigenvalue direction. Then, $\forall\,C\,,\gb\,,\beta$ and for all points along the fixed line, the real parts of the eigenvalues are always negative and ${\bf M}$ is stable. There exist two bifurcation points along ${\bf M}$:
\begin{equation}
 	z=\sqrt{\frac{3}{8C^2}}\,, \quad {\rm and} \quad z=\sqrt{\frac{9}{8\beta^2}}
	\label{bifurcationM}
\end{equation} 
Hence, for point ${\bf M}$ to be a stable node in the $\{y,z,r\}$ subspace, $z\leq\sqrt{3/8C^2}$, otherwise it is a stable spiral, whilst for point ${\bf M}$ to be a stable node in the $\{x,s\}$ subspace, $z\leq\sqrt{9/8\beta^2}$, otherwise it is a stable spiral.
\end{list}

% ----------------------------------------------------------------
\section{`Q'--Variable Autonomous System}\label{appdx:autonomousSystemQ}

With the compact variables defined by Eqs.~(\ref{eqn:Qvariables}), the axion--modulus system may be written:
\begin{align}
         x_Q' &= -\left[\frac{Q'}{Q} \pm H_Q \right]x_Q - Ms_Q\sqrt{z_Qt_Q\Phi} \,,  \nonumber \\
      	y_Q' &= -\left[\frac{Q'}{Q} \pm H_Q \right]y_Q \nonumber  \\ 
			&\qquad\qquad + \frac{C}{\sqrt{2}}\left[2z_Q^2 - z_Qt_QD\Phi+s_Q^2 \right] \,,  \nonumber \\
	z_Q' &= -\left[\frac{Q'}{Q} + \sqrt{2}Cy_Q \right]z_Q \,,  \nonumber \\
	s_Q' &= -\left[\frac{Q'}{Q} + \frac{1}{\sqrt{2}}Cy_Q\right]s_Q + Mx_Q\sqrt{z_Qt_Q\Phi} \,,  \nonumber \\
	t_Q' &= -\frac{Q'}{Q}t \,,
	\label{autonomousQ}	
\end{align}
with
\begin{eqnarray}
\frac{Q'}{Q} &=& -\frac{H_Q}{2}\bigg[ \gamma_{\rm b}(1 - x_Q^2 - y_Q^2 - z_Q^2 - s_Q^2 -t_Q^2)  \nonumber \\
	&+& 2x_Q^2 + 2y_Q^2\bigg] - \sqrt{\frac{1}{2}}z_Qt_Qy_Q\Phi\,,
	\label{QoQprime}	
\end{eqnarray}
and
\begin{equation}
H_Q=\pm\sqrt{3\left(1-z_Qt_QD\Phi\right)}\,,
	\label{HQ}
\end{equation}
where $\Phi=1/\sqrt{B\rho_\Lambda}$. Here, $(')=\frac{1}{Q}\frac{\rm d}{{\rm d}t}$. If the universe is expanding, the sign of the square root in Eq.~(\ref{HQ}) for $H_Q$ is positive. If the universe is contracting, the negative root should be chosen.\\

% ----------------------------------------------------------------
\section{Scanning the Initial Condition Manifold}\label{appdx:ICM}
The autonomous system~(\ref{eqn:autonomousSystem}) has eight different parameters which determine the subsequent motion of any given trajectory in phase space: six initial conditions, $\{ x_{i},y_{i},z_{i},r_i,s_{i},t_{i}\}$ and two parameters, $\{C,\beta\}$. To ensure that the entire space of initial conditions are sampled in an efficient around a region of interest, we use the method outlined in this appendix.\\

We begin by acknowledging the Friedman constraint
\be
\Omega_{\rm b}({\rm initial})=1-(x_i^2+y_i^2+z_i^2-r_i^2+s_i^2+t_i^2)\,, \nonumber \\
\ee
and the vacuum constraint $\zeta=\frac{t_i^2z_i^2}{r_i^4}$. Suppressing the subscript $i$ for brevity, the initial conditions are constrained to lie on the three--dimensional manifold $\mathcal{M}$:
\be
\label{appdix:manifold}
s^2=p-\zeta\frac{r^4}{z^2}+r^2-z^2\,,
\ee
where $p=1-\Omega_{\rm b}-x^2-y^2$. For given values of $x$ and $y$, the problem then reduces to varying two initial conditions evenly over $\mathcal{M}$, with the third constrained by Eq.~(\ref{appdix:manifold}). For example, if we were interested in scanning parameter space near to fixed point ${\bf A}$, we could, with only some loss of generality, make the simplifying assumption that the axion and modulus fields begin frozen, $x=y=0$. For trajectories starting near to fixed point ${\bf A}$, we have $p=1-\Omega_{\rm b}\approx0.01$. We can then vary $z$ and $r$, whilst still being free to independently vary $\{C,\beta\}$. If on the other hand, we were interested in scanning parameter space near to fixed point ${\bf G}$, since the fluid density and the value of $y$ at the critical point depends on $C$, $\Omega_{\rm b}=1-\frac{3\gb}{C^2}$, $y_c=\sqrt{\frac32\frac{\gb}{C}}$, we are not free to independently vary $C$ and $\{z\,,r\}$. Each time $C$ is changed, the equation for $\mathcal{M}$ also changes. This must be taken into account when using the method that is discussed below.

The value of $\zeta$ determines the topology of the initial condition manifold, which has non--constant curvature. If the vacuum is dS, $\zeta>\frac14$, the surface area of the manifold above some $s_{\rm min}$ is finite. For $\zeta\le\frac14$, (AdS and Minkowski vacua) the manifold is not bounded above $s_{\rm min}$ and its surface area is infinite. This change in topology is most easily illustrated by considering the intersection of the manifold $\mathcal{M}$ with the $s=s_{\rm min}$ plane. Solving Eq.~(\ref{appdix:manifold}) at $s=s_{\rm min}$ for $z$, generates four roots, two of which are the ones applicable for this problem:
\begin{align*}
		f_\pm&(r)=\frac{1}{\sqrt{2}}\bigg[p + r^2 - s_{\rm min}^2 - \\
	&\sqrt{p^2 + 2pr^2 - 2ps_{\rm min}^2 + r^4 - 2r^2s_{\rm min}^2 + s_{\rm min}^4 - 4\zeta r^4}\bigg]^{\frac12}\,,
\end{align*}
These two functions, $f_+(r)$ and $f_-(r)$ intersect at $r=r_{\rm max}$. Equating them and solving for $r$ gives the physically relevant root:
\be
\label{appdix:r_max}
r_{\rm max}=\sqrt{\frac{(p-s^2_{\rm min})(1+2\sqrt{\zeta})} {4\zeta-1} }\,.
\ee
Until these functions intersect at $r_{\rm max}$ we have $f_-(r)<f_+(r)$. Now, solving Eq.~(\ref{appdix:manifold}) at $s=s_{\rm min}$ for $r$ gives the two relevant roots:
\begin{equation}
	g_\pm(z)=\left[\pm\frac{z}{2\zeta}\left( \pm z+\sqrt{z^2+4\zeta(p- z^2 -s^2_{\rm min})}\right) \right]^{1/2}\,, \nonumber \\
	\label{appdix:r_roots}	
\end{equation}
which intersect at $z=z_{\rm max}$. Then, equating $g_+(z)$ and $g_-(z)$ and solving for $z$ gives the physically relevant root:
\be
\label{appdix:z_max}
z_{\rm max}=\sqrt{\frac{4\zeta(p-s^2_{\rm min})}{4\zeta-1}}\,.
\ee
Eqs.~(\ref{appdix:r_max}) and~(\ref{appdix:z_max}) illustrate the change in topology of $\mathcal{M}$: for $\zeta=1/4$, $r_{\rm max}\,,z_{\rm max}\rightarrow\infty$ and functions $f_\pm(r)$ and $g_\pm(z)$ and become parallel at large $r$ and $z$ and never meet; for $\zeta<1/4$, $r_{\rm max}\,,z_{\rm max}\in\mathbb{C}$ and the functions $f_\pm(r)$ and $g_\pm(z)$ diverge at large $r$ and $z$. In both cases, the manifold never intersects the $s=s_{\rm min}$ plane. Only for $\zeta>1/4$ is the surface area of the manifold bounded above $s_{\rm min}$. For models where the vacuum is dS, if an $s_{\rm min}$ is specified, the initial conditions on $z$ and $r$ may then be varied from some $r_{\rm min}\,,z_{\rm min}$ to the maximum values given by Eqs.~(\ref{appdix:r_max}) and~(\ref{appdix:z_max}). 

Since $\mathcal{M}$ has non--constant curvature, it is not trivial to sample the manifold in a uniform way. To guarantee approximately uniform coverage, we project a grid of $N$ squares,  each of which have the same area, $\mathcal{A}$, in a Cartesian coordinate system, onto $\mathcal{M}$. Since $\mathcal{M}$ has non--constant curvature, the projected shapes will be four--sided polygons of unequal area. From the induced metric, $\tilde{g}_{ab}$, on $\mathcal{M}$ we can compute the surface area of each polygon. The components of $\tilde{g}_{ab}$ read:
\begin{eqnarray*}
	\tilde{g}_{11} &=& 1 + \frac{1}{4s^2}\left(-\frac{4\zeta r^3}{z^2}+2r\right)^2\,, \nonumber \\
	\tilde{g}_{12} &=& \tilde{g}_{21}=\frac{1}{4s^2}\left(-\frac{4\zeta r^3}{z^2}+2r \right) \left( \frac{2\zeta r^4}{z^3}-2z\right)   \,,  \nonumber  \\
	\tilde{g}_{22} &=& 1 + \frac{1}{4s^2}\left(\frac{2\zeta r^4}{z^3}-2z \right)^2   \,.\nonumber 
	\label{eqn:metricomps}	
\end{eqnarray*}
The surface area $\tilde{\mathcal{A}}_i$ of each polygon $i$ is then
\be
\tilde{\mathcal{A}}_i = \mathlarger{\int\int}_{\mathcal{A}} \sqrt{\tilde{g}}\, {\rm d}r\,{\rm d}z\,, \nonumber  \\
\ee
where 
\begin{eqnarray*}
\tilde{g}&=&\text{det}\,[\tilde{g}_{ab}]=\tilde{g}_{11}\tilde{g}_{22}-\tilde{g}_{12}^2 \nonumber \\
	&=&\frac{\zeta^2 r^8+4\zeta^2 z^2r^6-7\zeta z^4r^4 +2z^6r^2 +z^6 p }{z^4\left(pz^2 -\zeta r^4 +z^2r^2 - z^4 \right) }\,.
\end{eqnarray*}
The limits of integration are the $\{r\,,z\}$ boundaries of each square in the Cartesian coordinate system. If these limits lie outside of the manifold, the boundary functions, $f_\pm(r)$, where $\mathcal{M}$ intersects $s_{\rm min}$ are used instead.

The initial condition manifold is now divided into polygons of different area. We randomly `throw' pairs of initial condition coordinates $\{r_i\,,z_i\}$ into each polygon on $\mathcal{M}$. The number of initial condition pairs that are thrown is proportional to the area of each polygon: the larger the polygon area, the more initial condition pairs are thrown. Since the total number of points thrown onto the manifold is large, this statistical method ensures an approximately even sampling of the initial condition manifold. 

% ---------------------- BIBLIOGRAPHY -----------------------------------------

\bibliographystyle{apsrev}
\bibliography{doddyoxford,ppxet}

\end{document}